# Speckle-Free Coherence Tomography of Turbid Media

**Authors:** Orly Liba[1,2,3,4], Matthew D. Lew[1], Elliott D. SoRelle[1,3,4,5], Rebecca Dutta[1,3,4], Debasish Sen[1,3,4], Darius M. Moshfeghi[6], Steven Chu[4,5,7], Adam de la Zerda[1,2,3,4,5]*

**Affiliations:**

[1]Department of Structural Biology, Stanford University, Stanford, CA 94305.

[2]Electrical Engineering, Stanford University, Stanford, CA 94305.

[3]Molecular Imaging Program at Stanford, Stanford, CA 94305.

[4]The Bio-X Program, Stanford, CA 94305.

[5]Biophysics Program at Stanford, Stanford, CA 94305.

[6]Byers Eye Institute, Dept. of Ophthalmology, Stanford University School of Medicine, Palo Alto, CA 94303.

[7]Departments of Physics and Molecular & Cellular Physiology, Stanford University, Stanford, CA 94305.

*Correspondence to: adlz@stanford.edu

**Abstract**

Optical coherence tomography (OCT) is a powerful biomedical imaging technology that relies on the coherent detection of backscattered light to image tissue morphology *in vivo*. As a consequence, OCT is susceptible to coherent noise (speckle noise), which imposes significant limitations on its diagnostic capabilities. Here we show a method based purely on light manipulation that is able to entirely remove the speckle noise originating from turbid samples without any compromise in resolution. We refer to this method as Speckle-Free OCT (SFOCT). Using SFOCT, we succeeded in revealing small structures that are otherwise hidden by speckle noise when using conventional OCT, including the inner stromal structure of a live mouse cornea, the fine structures inside the mouse pinna, sweat ducts, and Meissner's corpuscle in the human fingertip skin. SFOCT has the potential to markedly increase OCT's diagnostic capabilities of various human diseases by revealing minute features that correlate with early pathology.

**Main text**

**Introduction**

Since its initial demonstration nearly 25 years ago [1], optical coherence tomography (OCT) has become widely used by ophthalmologists for diagnosis of eye diseases [2]. Recently, OCT has gained popularity for its diagnostic capabilities in cardiology [3], dermatology [4–6], dentistry [7] and cancer research [8–10]. Because of the nature of imaging with coherent light, OCT suffers from speckle noise [11] that effectively causes significant degradation in spatial resolution and prevents the imaging technique from reaching a greater diagnostic potential. Speckle noise is inherent to all coherent imaging systems and arises from the interference of light scattered from multiple points within a turbid sample [12], such as biological tissue. Since the development of OCT, researchers have described various methods for decreasing speckle noise. One group of methods involves incoherently averaging (compounding) several images with uncorrelated speckle noise. A basic limitation of these compounding methods is that increased compromise in resolution or depth of field is required to further decrease speckle noise. Hence, these methods can never eliminate speckle noise entirely. Obtaining images with non-correlated speckle patterns can be achieved by various methods, including: scanning from different angles, scanning several adjacent images, scanning with bands of different incident wavelengths, and scanning with different polarizations. These methods are referred to as angular [13], spatial [14], frequency [15] and polarization compounding [11], respectively. The second group of methods to reduce speckle noise is based on image processing techniques such as adaptive filters [16] and wavelet analysis [17], among others [18–20]. These methods cannot reveal information that was lost due to speckle and merely reduce the appearance of speckle noise. Other techniques such as speckle reduction by a partially spatially-coherent source [21] have been suggested in the past for OCT imaging, however, to date, they have not demonstrated speckle reduction in tomograms of turbid media.

In contrast, the technique we present here, Speckle-Free OCT (SFOCT), is capable of eliminating speckle noise entirely without compromising the resolution of the image. Hence, SFOCT clarifies and reveals structures that are otherwise obscured or undetectable. The fundamental concept of SFOCT is the introduction of time-variant local phase shifts within the light beam illuminating the sample. These variations translate into local phase shifts in the light reflected from scatterers within each voxel, which subsequently yield non-correlated speckle patterns that can be temporally averaged to create an image with reduced speckle noise (Fig. 1a). Since each image is acquired at the same angle, sample position, and set of illumination wavelengths, increasing the number of compounded images does not lead to an inherent degradation in resolution or depth of field. Because increasing the number of uncorrelated images does not reduce resolution, it is possible to average many images together and subsequently reduce speckle noise such that it is undetectable relative to other noise sources in the image.

An approximate mathematical description of this phenomenon is given by Eq. 1:

$$I = \frac{1}{M} \sum_{m=1}^{M} \left| \sum_{n=1}^{N} a_n e^{i\varphi_n} e^{i\theta_{n,m}} \right| \qquad (1)$$

in which $I$ is the pixel value after averaging $M$ images obtained at different times and with different local phase shifts within the illuminating beam. $N$ is the number of scatterers inside a voxel. For each scatter $n$ within that voxel, $a_n$ is the scattering amplitude (proportional to its amplitude reflection coefficient) and $\varphi_n$ is the phase delay due to the axial location of the scatterer. $\theta_{n,m}$ is the local phase shift of the illumination beam at the location of scatterer $n$, which changes in time in SFOCT. Simulations show how this approach decreases speckle noise in Fig. S1a.

**Results**

The implementation of SFOCT is straightforward and does not require specialized equipment or light sources. In fact, we describe here a method to adapt any OCT system into a SFOCT system. We have demonstrated SFOCT on two commercial spectral domain OCT (SDOCT) systems: a high-resolution (HR-OCT) skin-imager (Ganymede HR, Thorlabs) and a Food and Drug Administration (FDA)-approved retinal imager (iFusion, Optovue). For both devices, we implemented local, random time-varying phase shifts by translating a ground glass diffuser at an OCT conjugate image plane (Fig. 1b, Fig. S2). In HR-OCT, the diffuser was placed at the original focal plane of the OCT probe, and a new focal plane was projected by a 4f imaging system. Dispersion compensation elements were added to an extended reference arm to account for the addition of lenses. In the retinal OCT system, it was not necessary to project a new focal plane, because one such plane is accessible inside the original OCT probe (Fig. S2b). Hence, the retinal implementation of SFOCT is simpler, as the change in the sample arm is negligible and therefore does not require extension of the reference arm. In both systems, the diffuser is moved in a plane perpendicular to the optical axis by a motor. The image is acquired several times, imaging the same exact location of the sample but through different locations on the diffuser. The random time-varying pattern of the diffuser changes the speckle pattern of the image. After averaging $M$ measurements, speckle noise decreases by a factor of $\sqrt{M}$ [11]; for example, a mere 9 averages will lead to a 3–fold reduction in speckle noise.

We characterized the thickness profiles (Fig. 1c-d, Fig. S3) of the diffusers used in this study, as well as their effects on the power on the sample (Fig. S4a, Table S2), signal intensity (Fig. S4b, Table S3), and lateral resolution (Fig. S5, Table S4). Detailed acquisition parameters for all presented images are provided in Table S1.

A key advantage of SFOCT is its ability to improve effective resolution so that closely-spaced scattering objects can be distinguished. We demonstrated and quantified this improvement by imaging a small gap in a phantom made of titanium dioxide ($TiO_2$) powder dispersed in polydimethylsiloxane (PDMS). A narrowing gap was created by adjoining two rectangular pieces of the phantom at an angle (Fig. 1e, S6a). Using SFOCT, we were able to detect a gap that was 2.5 times smaller than the smallest gap detected with OCT, as determined by image segmentation (Fig. 1f-j, S6b-h, S7) [22]. Measurements of a resolution test target (Fig. S5, Table S4) showed that the smallest resolvable separation in a non-turbid sample (7.5 µm in OCT and 12.2 µm in SFOCT) was smaller than the smallest gap measured on the phantom (31.5 µm in OCT and 14 µm in SFOCT). These results prove that speckle noise effectively limits the resolution in most OCT images and that SFOCT is able to recover the loss in resolution. OCT and SFOCT images in this study are depicted on a logarithmic scale with black and white representing low and high signal intensities, respectively.

OCT speckle noise follows a Rayleigh distribution [15]. Strong evidence that SFOCT reduces speckle would be indicated by a change in pixel value statistics from a Rayleigh distribution toward the expected distribution of scatterers randomly dispersed within a phantom (a Poisson distribution) [22,23]. We experimentally demonstrated this change in statistics by measuring the pixel value distribution in a phantom made of large gold nanorods (LGNRs) [24] dispersed in an agarose gel (Fig. S8). Owing to the strong backscattering and high concentration of the metallic nanoparticles, the agarose-LGNR phantom is an excellent model for turbid media, and it produced Rayleigh speckle statistics, as expected for conventional OCT imaging (Fig. 2a). In contrast, the pixel value distribution obtained with SFOCT resembled a Poisson distribution, in which each scattering event contributes a signal value that is equal to the backscattering of a single LGNR (Fig. 2B, S9a). As predicted by Eq. 1, increasing the number of averages reduced speckle noise, reduced the broadness of the distribution of pixel values, and led to an SFOCT image that closely mimicked the random distribution of the scatterers in the phantom. Because the phantoms were composed of a random spatial distribution of discrete particles, we expected the speckle-free image of the phantom to be non-uniform (Fig. S1b).

To further validate that SFOCT removes speckle noise, we compared experimental data to the theoretical decrease of speckle contrast, proportional to $1/\sqrt{M}$, where $M$ is the number of compounded images with uncorrelated speckle noise [11]. Including the inherent signal variation of the sample, the decrease in normalized standard deviation (SD), $C$, can be described by:

$$C = \frac{\sigma}{\langle I \rangle} \tag{2a}$$

$$C^2 \langle I \rangle^2 = \sigma^2 = \sigma_0^2 + \sigma_{speckle}^2 \tag{2b}$$

$$\sigma_{speckle}^2 = \frac{\sigma_{speckle,0}^2}{M} \tag{2c}$$

Here, $\sigma$ is the measured SD in a region of interest (ROI), and $\langle I \rangle$ the average pixel intensity in the same ROI. $\sigma^2$ is the measured variance in pixel values, $\sigma_0^2$ is the intrinsic variation in signal (in this case, caused by the variation in number of particles in a voxel), $\sigma_{speckle}^2$ is the variance of the speckle noise, which decreased by a factor of $M$ during compounding, and $\sigma_{speckle,0}^2$ is the variance of the speckle noise without any compounding.

Conventional OCT images exhibited negligible reduction in normalized SD ($C$) even with extensive averaging ($M$=100), indicating that speckle noise was not affected by averaging. By comparison, SFOCT imaging with equivalent averaging led to a significant reduction in the normalized SD as a result of the reduction in speckle noise (Fig. 2c, S9b). The values for $\sigma_0^2$ and $\sigma_{speckle}^2$ were found using a non-linear least squares fit to equation (2b).

Since the normalized SD is composed of both speckle noise and the intrinsic random distribution of particles in the phantom (represented by $\sigma_0$), we define the normalized speckle (Eq. 3), which decreases by a factor of $\sqrt{M}$ theoretically (predicted by Eq. 2) and experimentally (Fig. 2d, S9c).

$$\text{Normalized speckle} = \sqrt{\frac{\sigma^2 - \sigma_0^2}{\sigma_{speckle,0}^2}} = \frac{1}{\sqrt{M}} \qquad (3)$$

The speckle reduction achieved with SFOCT reveals fine structures that are typically obscured by noise. As a practical demonstration of this ability, we embedded LGNRs and polystyrene beads of 3 µm diameter inside an agarose phantom (Fig. 2e). As predicted, speckle noise was predominant in conventional OCT images and consequently most of the beads were not visible (Fig. 2f, h and i). On the other hand, SFOCT imaging enabled detection of the beads along with the distribution of the LGNRs in the agarose phantom (Fig. 2g j, k). The evolution of the images as the number of averages increases (Fig. 2i, k) showed that the beads were more easily detected in the SFOCT image compared to OCT after as few as 10 averages. As the number of averages increased, the beads became more visible in SFOCT, while the OCT image remained overshadowed by an unchanged speckle pattern. Note that when the number of averaged images was low, photon shot noise was significant. The signal intensity profiles (Fig. 2m-n) show the reduction of speckle noise and the presence of the beads identified using SFOCT, while in the OCT profiles some of the beads were not visible or were close to the intensity of speckle noise. Images of the phantom acquired with bright field microscopy (Fig. S10a) and SFOCT revealed a sparse distribution of beads inside the agarose-LGNR phantom. This comparison indicated that SFOCT yielded a more accurate representation of the structure of the sample than OCT. Imaging an

agarose-TiO$_2$ nanopowder phantom with TiO$_2$ aggregates in it further validated the capability of SFOCT to produce images that better represent the true structure of the sample (Fig. S10b-d).

**Results in living intact tissue**

One of the greatest biomedical advantages of OCT is its ability to provide non-invasive high-resolution images of intact living tissues. However, strong speckle artifacts drastically limit the ability to resolve fine anatomical structures. These limitations become obvious upon comparison of OCT images with histological tissue sections. By removing the significant contribution of speckle noise, SFOCT is capable of rendering *in vivo* images that approach histological detail.

Figure 3 depicts OCT and SFOCT images of a mouse ear pinna, which consists of well-defined epithelial and cartilage layers, small blood and lymph vessels, and numerous hair follicles and sebaceous glands. Many of these structures were masked by speckle noise in OCT, but became visible in SFOCT images. Speckle removal allowed imaging of fine structures in cross-sectional B-scans (Fig. 3a-f) as well as in frontal (*en face)* sections (Fig. 3g-h), indicating that SFOCT provided major improvement in image quality in all three spatial dimensions. The arrows in Fig. 3d show an anatomical feature the size of 9.1 µm. Features this small are not observed in the OCT images due to speckle noise. The arrow in Fig. 3f shows a dark horizontal line which is 2 μm thick, showing that the intrinsic axial resolution, as defined by the broadness of the OCT spectrum, is uncompromised. The *en face* SFOCT image shows lymph vessels and fine structures that are more visible compared to the OCT image (Fig. 3g-h). Fig. 3i depicts a histological section of a pinna, which shows the small structures that were also observed clearly in SFOCT images but not in OCT images. We further compared SFOCT images with alternative speckle reduction methods including spatial compounding (Fig. S11), adaptive Wiener filtering (Fig. S12c-d), hybrid median filtering (Fig. S13a-c) and symmetric nearest-neighbor filtering (Fig. S13d-f). SFOCT outperformed each of these methods in terms of speckle noise reduction and provided highly detailed images, while in the alternative approaches, noise suppression came at a cost of smoothing fine features in the image.

The most common clinical application of OCT is in ophthalmic imaging. As a second demonstration of SFOCT, we acquired images of the cornea and retina of a live mouse. Using SFOCT, we were able to see the lamellar structure of the corneal stroma as well as clear boundaries between various layers of the cornea (Fig. 4a-e). Due to speckle noise, conventional OCT was neither able to show clear boundaries between layers nor show the structure of the stroma. We then imaged the retina of a live mouse using a similar imaging setup [22]. The individual layers of the retina were particularly well-resolved with SFOCT (Fig. 4f-i). For example, the outer plexiform layer and the external limiting membrane can be readily distinguished in SFOCT images.

To demonstrate the potential of SFOCT in dermatological applications, we imaged the fingertip skin of a human volunteer (Fig. 5, S14, and S15). The speckle noise reduction achieved with SFOCT enabled detection of fine structures within the fingertip including sweat ducts and tactile corpuscles. To our knowledge this is the first time that the tactile corpuscle has been clearly observed in the intact skin of a live human. SFOCT was particularly helpful in identifying the boundaries between the corpuscle and the surrounding dermis. As in images of the mouse cornea, SFOCT images of the fingertip revealed the cellular structure and striation of the tactile corpuscle, proving that SFOCT can remove speckle noise without compromising resolution. This example suggests that SFOCT may be used to improve non-invasive dermatological studies in humans by producing images that approach the quality of histology.

We also performed SFOCT retinal imaging of a human volunteer (one of the authors). Fig. S16 depicts images of the human retina obtained with SFOCT using the retinal system described in Fig. S2. As in mouse retinal imaging studies, these images show clearer differentiation between retinal layers. Optical removal of speckle resulted in a greater delineation of the various retina layers, seen most clearly in all three nuclear layers as well as in the differentiation between the outer retina layers.

In order to allow SFOCT imaging of moving samples, we implemented our approach using A-scan averages instead of frame averages. This requires moving the diffuser fast enough such that it is translated by approximately one wavelength between A-scan acquisitions, resulting in uncorrelated speckle patterns, which form a speckle-free A-scan when averaged. The diffuser was moved rapidly and continuously using a rotating mount (Pacific Laser Equipment). Scanning was near the edge of the rotating diffuser where the velocity was 9 mm/s. We used this setup to image the cornea of a mouse *in vivo* (Fig. S17).

**Discussion**

We have demonstrated SFOCT, a technique that is able to efficiently eliminate speckle noise in OCT by utilizing a moving diffuser to locally induce random phase shifts in the light illuminating and collected from within each voxel. The method we have shown is very effective, as well as low cost, robust, and easily adaptable to existing OCT systems. In this study, SFOCT was integrated as an extension to two commercial OCTs with basic components. Our implementation utilized a simple ground glass moving diffuser to reduce speckle, however the same physical concept can likely be created by other means, for example by a spatial light modulator as long as the phase changes within the voxel. The ability to scramble the phase inside the voxel is limited by the point spread function (PSF) of the lenses in the 4f imaging system. Therefore, these lenses should have a smaller PSF compared to the main lens of the OCT, which defines the OCT's lateral resolution and voxel size. This means that the size of the voxel of the OCT has to be deliberately larger in order to introduce phase scrambling within it. The benefit of speckle noise removal significantly outweighs the reduction in lateral resolution because it ultimately allows detection of fine detail.

The use of a moving diffuser placed in the optical path in order to reduce speckle noise has been previously demonstrated for imaging and display [12,25–27] and the effects of partially coherent illumination on reducing the speckle artifacts present in coherent imaging have been previously explored [28,29]. However, this study presents the first implementation of a moving diffuser in OCT and the first demonstration of speckle-free high-resolution tomograms of turbid media captured with coherent illumination. Recently, random laser illumination [30] and a low-spatial-coherence semiconductor laser [31] have been proposed speckle-free light sources for imaging, but they have not been demonstrated to produce tomograms. Furthermore, SFOCT is able to eliminate speckle originating from the turbid sample itself (caused by multiple back scattering from the imaged voxel) in addition to speckle caused by a turbid object placed in the optical path (caused by multiple forward scattering).

One theoretical limitation of our method is that an object cannot move more than a few microns while frame averages are acquired. However, this requirement is merely an artifact of acquiring frame (B-scan) averages instead of A-scan averages, which were demonstrated in Fig. S17 by using a rotating diffuser. Another way to acquire SFOCT images of fast-moving objects is by implementing a conventional tissue-tracking system[32]. Overall, we do not expect the averaging requirement to limit SFOCT imaging because significant speckle reduction can be achieved with as few as 10 averages, an amount already used in conventional OCT to reduce photon and thermal noise. Further, we demonstrate here that despite the increase in acquisition time, it is possible to image living subjects' skin and eyes. As hardware advances continue to improve OCT acquisition rates, SFOCT acquisition times will also improve. Moreover, researchers have developed OCT systems that achieve image compounding without extending the acquisition time, such as interleaved OCT[33], which could be applied to SFOCT as well. A detailed discussion of the effects of SFOCT on resolution, signal intensity, and acquisition time appears in the supplementary discussion [22].

In summary, we expect that SFOCT will enable superior diagnostic capabilities compared to conventional OCT because of its ability to reveal anatomical features that are otherwise hidden by speckle noise. Potential clinical applications of SFOCT include early detection of epithelial cancers, evaluation of tumor margins, and early detection of retinal diseases such as atrophic and neovascular age-related macular degeneration. Additionally, the elimination of speckle noise facilitates further OCT image enhancement including blur-deconvolution for extended depth of field, super resolution, and improved segmentation of structures such as retinal layers [34,35], which will aid the diagnosis of diseases.

**Methods**

For detailed methods see the *supplementary section*[22].

**Experimental setup.** SFOCT was implemented by modifying two existing OCT systems: the Ganymede HR (Thorlabs) and a clinical retinal imaging device (iFusion , Optovue). Both are SDOCT systems. In the Ganymede HR ($\lambda$=800-1000 nm), the diffuser was placed at the original focal plane of the OCT probe, and a new focal plane was projected by a 4f imaging system using two similar lenses (LSM02-BB, Thorlabs). Due to the extension of the sample arm and the addition of 2 lenses and the diffuser, the reference arm was extended by approximately 10 cm and two dispersion compensation elements were added (LSM02DC, Thorlabs). The diffuser was placed in the focal plane of the first lens and held within a custom motorized mount with XYZ translation (based on CXYZ1, Thorlabs). The diffusers were moved by a motor (Z812, Thorlabs), back and forth along one axis, and controlled through computer software (Thorlabs APT, Thorlabs). Post processing was done with Matlab (Mathworks). The implementation of SFOCT on the iFusion appears in Fig. S2. The diffuser was placed in the conjugate image plane (Fig. S2b). The diffuser was held within a thin fixed mount (LH-1T, Newport) which was attached to a motorized translation stage. All processing was done internally by the iFusion computer and software. The diffusers used for all experiments are ground glass diffusers with anti-reflective coating on one side (DG10-1500-B and DG10-2000-B, Thorlabs). The 1500 and 2000 grit diffusers are 2 and 1 millimeter thick, respectively. The 3 µm lapped diffuser was created by further lapping a commercial 1500 grit diffuser with 3 µm aluminum oxide grit (Universal Photonics) for 15 minutes. The profile and height statistics of the diffusers appear in Fig. 1c-d and Fig. S3.

**Processing and display.** The number of averages for each image appears in Table S1. The averaged image is displayed on a logarithmic scale with image-adaptive brightness scaling unless otherwise stated. Dark pixels correspond to low scattering from the sample while bright pixels correspond to high intensity of scattering.

**Imaging of live mouse pinna, retina and cornea.** All animal experiments were performed in compliance with IACUC guidelines and with the Stanford University Animal Studies Committee's Guidelines for the Care and Use of Research Animals. Experimental protocols were approved by Stanford University's Animal Studies Committee. Nu/nu mice (Charles River Labs) were anesthetized by inhalation of 2.5% isoflurane in $O_2$ (v/v). Once adequately anesthetized, the right ear pinna was immobilized using double-sided tape. We optimized light transmission to the sample by applying ultrasound (US) gel to the mouse skin and covering the gel with a 2 millimeter thick one sided anti-reflective coated glass (650-1050 nm), with the coating at the air-glass interface. For retinal and corneal imaging studies, mice were anesthetized with intraperitoneal injections of 80 mg/kg ketamine (Vedco Inc.) and 10 mg/kg xylazine (Lloyd Inc.). Once adequately anesthetized, the mice were mounted on to a platform and secured with a stereotactic device. With the mouse secured, the stage was tilted to orient the mouse's eye upward, with the top of the cornea being approximately parallel to the table. Pupillary dilation was achieved by applying 1 drop each of 1% tropicamide (Bausch & Lomb), and 2.5% phenylephrine hydrochloride (Paragon BioTeck) to the eyes, for 2 min each. 2.5% hypromellose solution (Gonak™, Akorn Inc.) was then placed over each eye as a contact solution. Anesthesia was continually maintained using a nose-cone delivering 1.5-2% isofluorane in $O_2$.

**Imaging of human fingertip.** The fingertips of healthy volunteers were imaged with the Ganymede HR. The subject's finger was pressed onto the bottom of a fixed glass window with anti-reflective coating at the air-glass interface. As in the setup for imaging the mouse pinna, we optimized light transmission to the sample by applying US gel to the fingertip skin.

**Imaging of human retina.** A healthy human volunteer (one of the authors) was scanned with the retinal system (iFusion, Optovue). Stanford Office of Human Subjects Research determined that this did not warrant for an institutional review board. The diffuser was placed in the image plane for SFOCT imaging and removed for OCT imaging. Images were acquired quickly one after the other to reduce movement between scans.

**Supplementary Materials:**

Materials and Methods

Figures S1-S17

Tables S1-S4

**Acknowledgements:**

This work was funded in part by grants from the Claire Giannini Fund, the United States Air Force (FA9550-15-1-0007), the National Institutes of Health (NIH DP50D012179), the National Science Foundation (NSF 1438340), the Damon Runyon Cancer Research Foundation (DFS# 06-13), the Susan G. Komen Breast Cancer Foundation (SAB15-00003), the Mary Kay Foundation (017-14), the Donald

E. and Delia B. Baxter Foundation, the Skippy Frank Foundation, the Center for Cancer Nanotechnology Excellence and Translation (CCNE-T; NIH-NCI U54CA151459), and the Stanford Bio-X Interdisciplinary Initiative Program (IIP6-43). A. dlZ. is a Pew-Stewart Scholar for Cancer Research supported by The Pew Charitable Trusts and The Alexander and Margaret Stewart Trust. O.L. is grateful for a Stanford Bowes Bio-X Graduate Fellowship. E.D.S. wishes to acknowledge funding from the Stanford Biophysics Program training grant (T32 GM-08294). We wish to thank Dr. Joseph Kahn for insightful discussions, Roopa Dalal for images of tissue sections, Ayana Henderson for useful discussions, Timothy R. Brand and the Ginzton Crystal Shop for creating the lapped diffuser, Stanford Neuroscience Microscopy Service (supported by NIH NS069375), and Jim Strommer for custom artwork in figure 1. We appreciate the help of Dr. Audrey (Ellerbee) Bowden and her lab, especially Gennifer Smith, for her help with creating phantoms.

**Author contributions:**

O.L., M.D.L., E.D.S., S.C. and A.dlZ. conceived and designed the research; O.L., M.D.L., E.D.S., R.D. and D.S. performed experiments; D.M.M. contributed tools and expert advice; O.L. analyzed data; O.L., M.D.L., E.D.S., R.D., D.S., D.M.M. and A.dlZ. co-wrote the paper. All authors have given their approval for this work.

**Competing financial interests:**

O.L, M.D.L, E.D.S and A.dlZ. are listed as inventors on a USPTO provisional patent application (62/243466) related to the work presented in this manuscript. All other authors have nothing to disclose.

Correspondence and requests for materials should be addressed to adlz@stanford.edu

**Figures:**

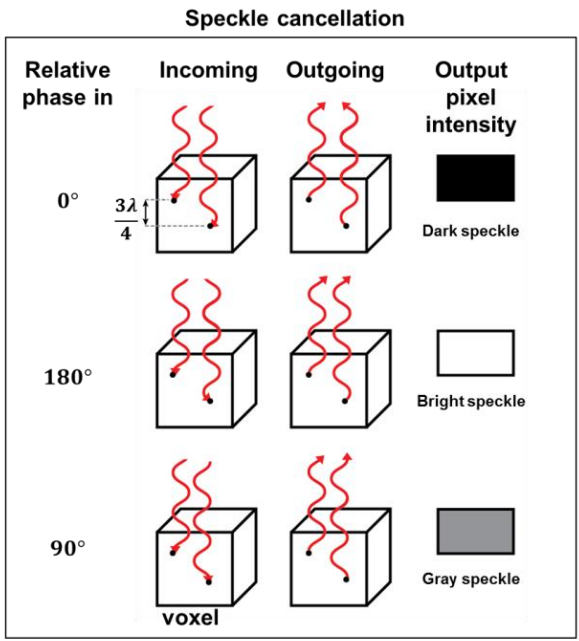
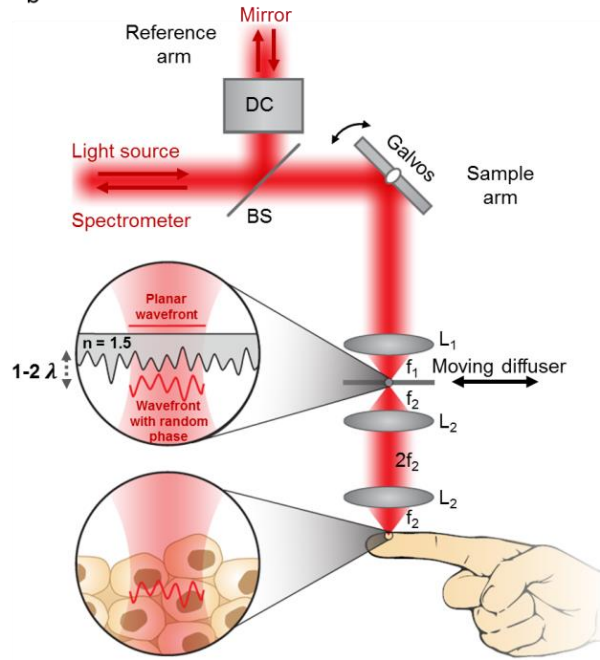
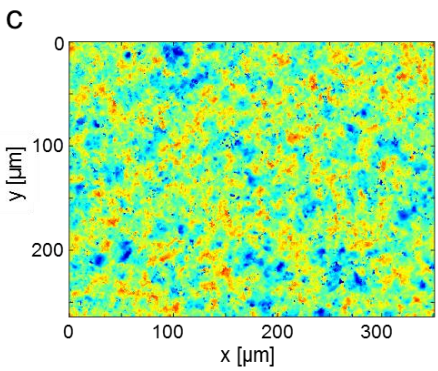
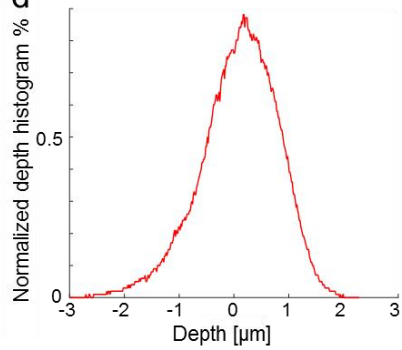
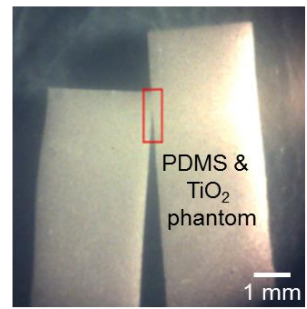
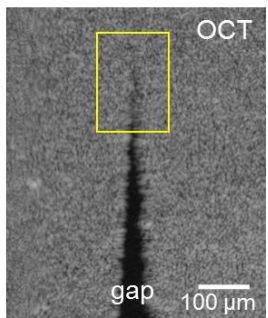
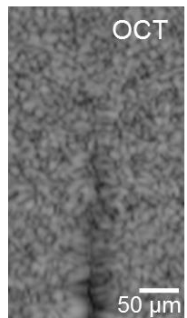
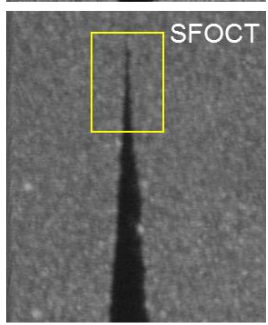
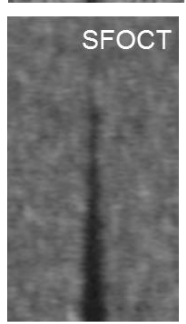
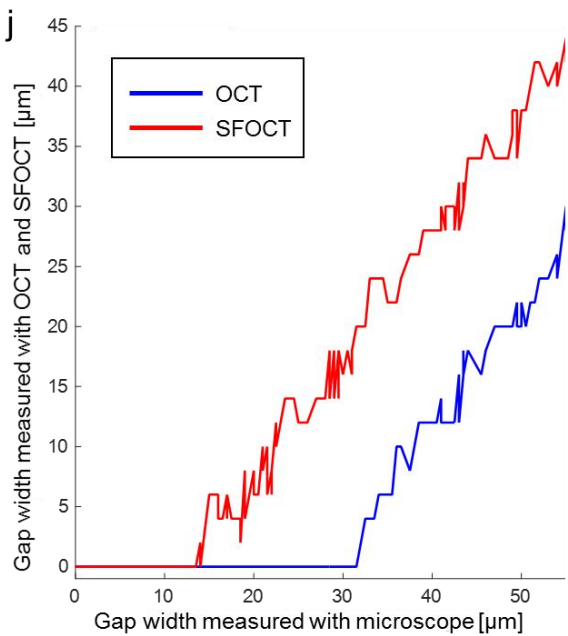

**Figure 1 | SFOCT speckle removal concept and demonstration of improved spatial resolution. a,** Introducing local phase shifts between scatterers in the same voxel changes the intensity of the resulting speckle noise, enabling one to reduce speckle noise via averaging many different phase shifts. This leads to the detection of scatterers otherwise hidden by the speckle noise. **b,** Implementation of SFOCT on the HR-OCT system. DC, dispersion compensation; BS, beam splitter; $L_1$, lens of the conventional OCT; $L_2$, lenses added to create a 4f imaging system; $f_1$, focal length of $L_1$; $f_2$, focal length of $L_2$; n, refractive index of the diffuser; $\lambda$, the center wavelength of the light source. **c,** Depth profile of the 1500 grit diffuser measured with a non-contact 3D optical profiler. **d,** Depth histogram of the 1500 grit diffuser. **e**, A phantom composed of PDMS and TiO2 powder was shaped to form a gap of decreasing size to evaluate the spatial resolution of SFOCT versus OCT. **f, g,** OCT and SFOCT *en face* scans inside the phantom. **h, i,** Close-up view on the regions marked in **f** and **g** showing the micron-size gap. The gap which is clearly visible in SFOCT does not appear in the OCT image due to speckle noise. **j,** The size of the gap measured from the OCT and SFOCT images versus the size of the gap measured from a bright-field microscope image (10x, NA=0.25). The minimum size of a resolvable gap is decreased by a factor of 2.5 owing to SFOCT. Note that the visibility of the gap is limited only by the speckle created by the turbid PDMS-$TiO_2$ phantom.

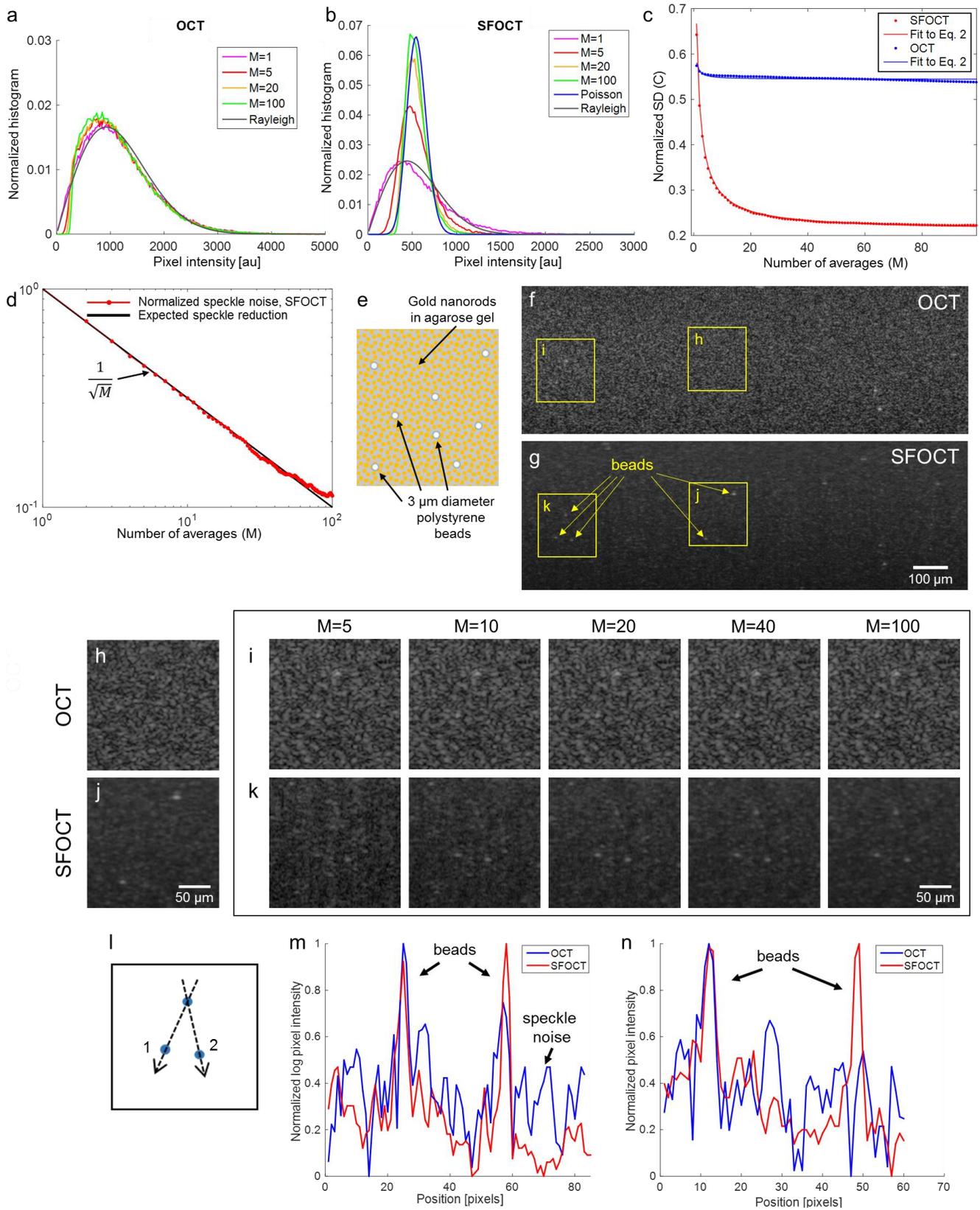

**Figure 2 | Demonstration of speckle removal and increased visibility of SFOCT in phantoms. a, b,** Statistical analysis of pixel values of scans of an agarose-LGNR phantom obtained with OCT and SFOCT. The OCT image is dominated by speckle noise and the distribution of pixel values is approximately a Rayleigh distribution that

persists with averaging (M is number of averages). In the SFOCT image, increasing the number of averages moves the distribution of pixel values towards a Poisson distribution, which expresses the probability of a given number of LGNRs to be present in a single voxel. **c,** Reduction in normalized standard deviation (SD) versus the number of averages, $M$, for OCT and SFOCT. The reduction in the normalized SD is significantly larger in SFOCT versus OCT. **d**, The reduction of normalized speckle as defined by equation (3) follows $1/\sqrt{M}$, as expected. **e,** Schematic of a phantom made by dispersing LGNRs and 3 µm diameter polystyrene beads in an agarose gel. **f, g,** OCT and SFOCT B-scans of the phantom. In the OCT image, many of the beads cannot be detected due to speckle noise. In the SFOCT image, speckle noise is eliminated and the beads become visible, along with the random distribution of LGNRs in the phantom. **h, j**, **i, k** close-up views of regions in the phantom showing superiority of SFOCT over OCT in detecting the beads. In the SFOCT image the beads are revealed as the number of averaged images (M) increases. **l,** Schematic showing the locations of the three beads. **m,** Intensity profiles (on logarithmic scale) along line 1 depicted in **l**. **n,** Intensity profiles along line 2 depicted in **l** demonstrating the beads are easily visible in SFOCT but not in OCT.

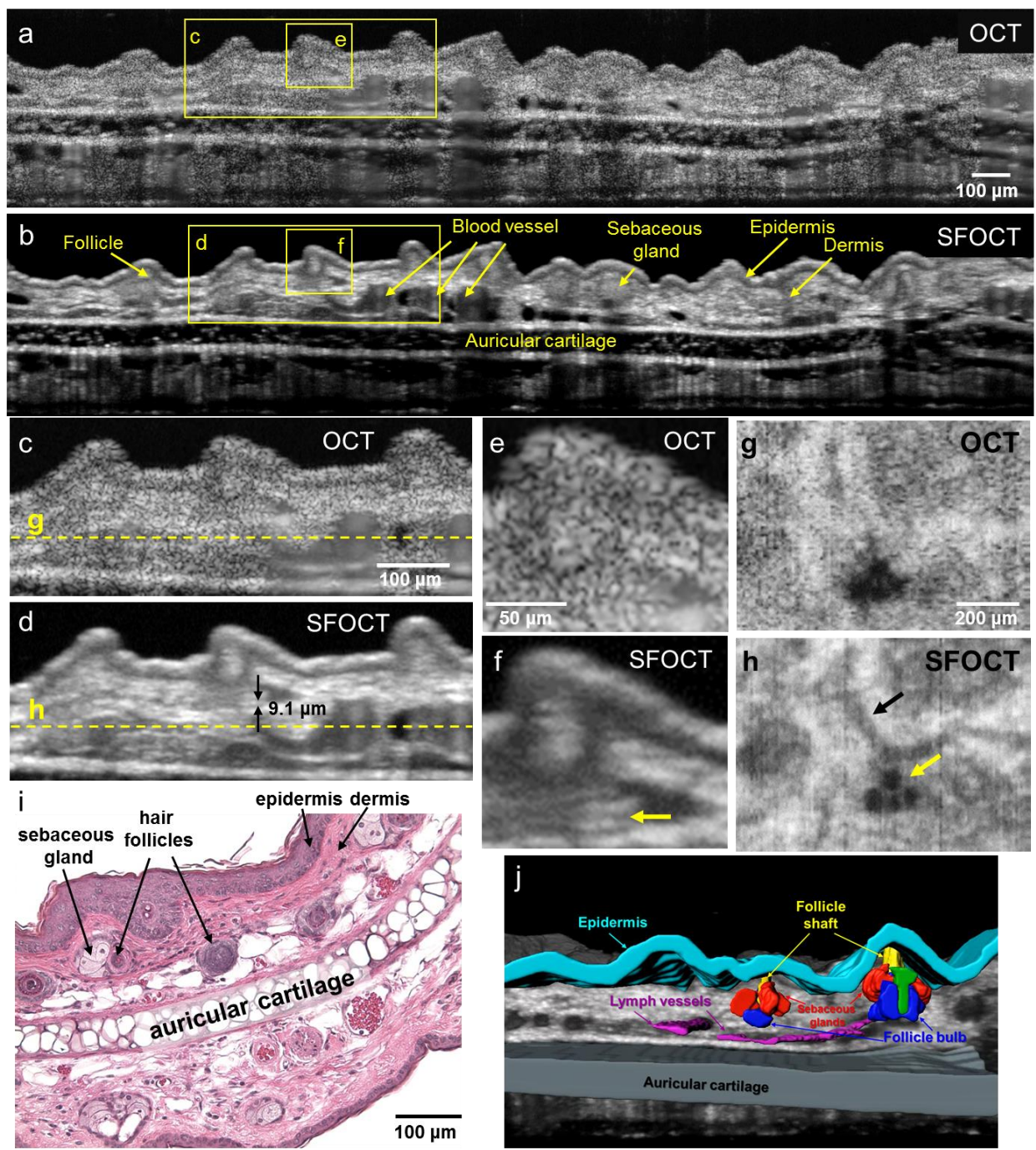

**Figure 3 | SFOCT imaging of the mouse ear pinna reveals fine low-contrast structures such as lymph vessels, hair follicles and sebaceous glands. a,** OCT B-scan of a mouse pinna. **b,** SFOCT scan of a mouse pinna. **c, d,** Close-up views on the regions marked in **a, b**. **e, f,** Close-up views on the regions marked in **c**. The arrows in **d** depicts an anatomical feature the size of 9.1 µm that is not visible in the conventional OCT image. The arrow in **f** shows a dark line which is 2 µm thick. **g, h,** OCT and SFOCT *en face* scans at the depth indicated by the dashed line in **c, d**. The SFOCT image in **h** shows lymph vessels (black arrow) and fine structures (yellow arrow) that are nearly invisible in **g**. **i,** Microscope image of an H&E stained mouse ear pinna at 10x magnification. **j,** Segmentation of the ear volume is possible owing to the removal of speckle noise, revealing the structure of hair follicles. Cyan-epidermis, gray- auricular cartilage, magenta- lymph vessels, red- sebaceous glands, blue- follicle bulb, yellow- follicle shaft, green- unidentified part of the follicle.

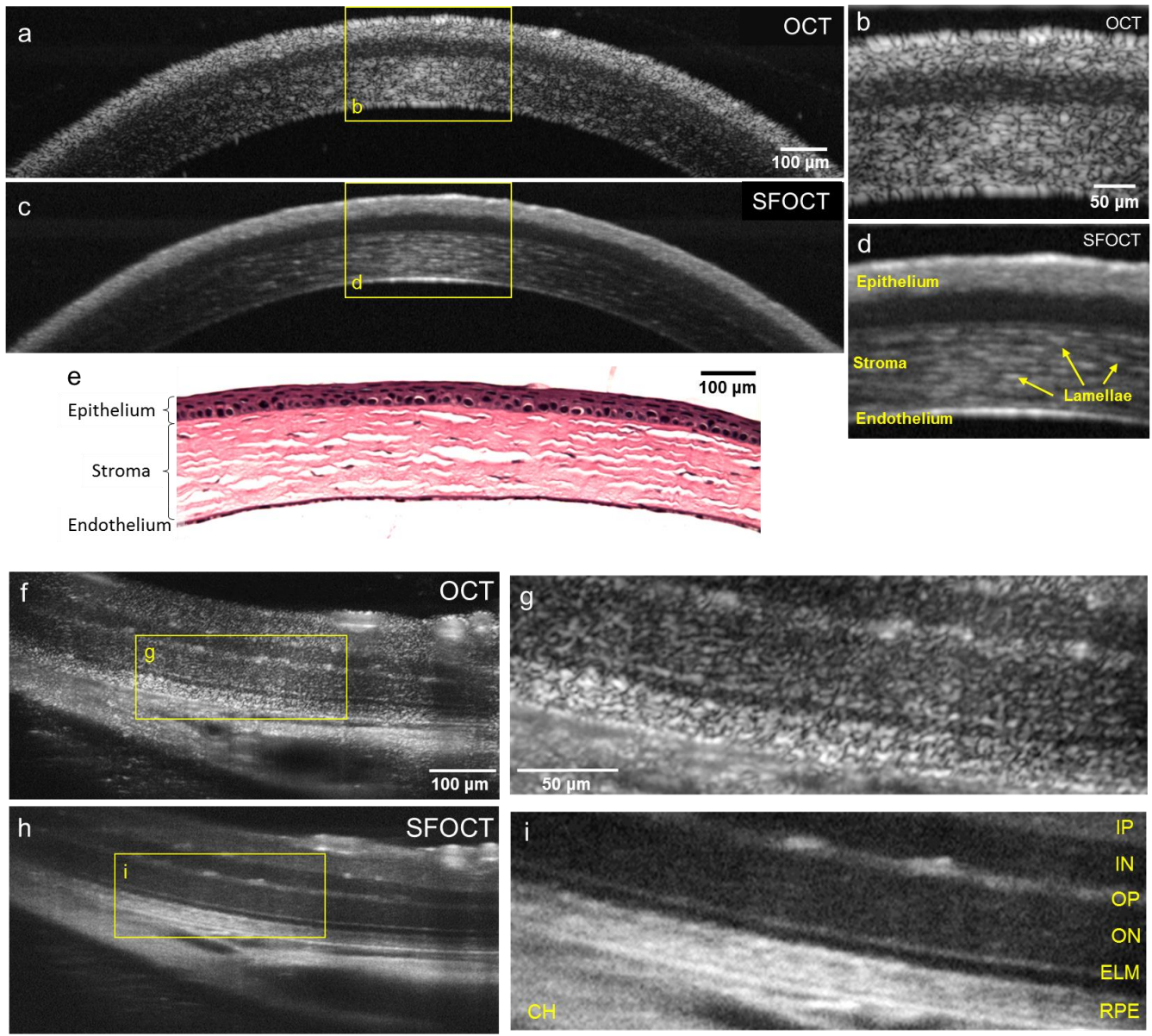

**Figure 4 | SFOCT imaging of the mouse cornea and retina clarifies the boundaries between the layers and reveals the cellular structure of the stroma. a,** OCT B-scan of a mouse cornea. **b,** Close-up view on the region marked in **a**. Strong speckle noise, due to the high density of scatterers in this tissue, is masking the inner structure of the stroma. **c,** SFOCT scan of a mouse cornea. **d,** Close-up view on the region marked in **c**. **e,** Microscope image of H&E stained mouse cornea at 10x magnification. **f,** OCT B-scan of a mouse retina. **g,** Close-up view on the region marked in **f**. **h,** SFOCT scan of a mouse retina. **i,** Close-up view on the region marked in **h**. IP, inner plexiform; IN, inner nuclear layer; OP, outer plexiform layer; ON, outer nuclear layer; ELM, external limiting membrane; RPE, retinal pigment epithelium; CH, choroid.

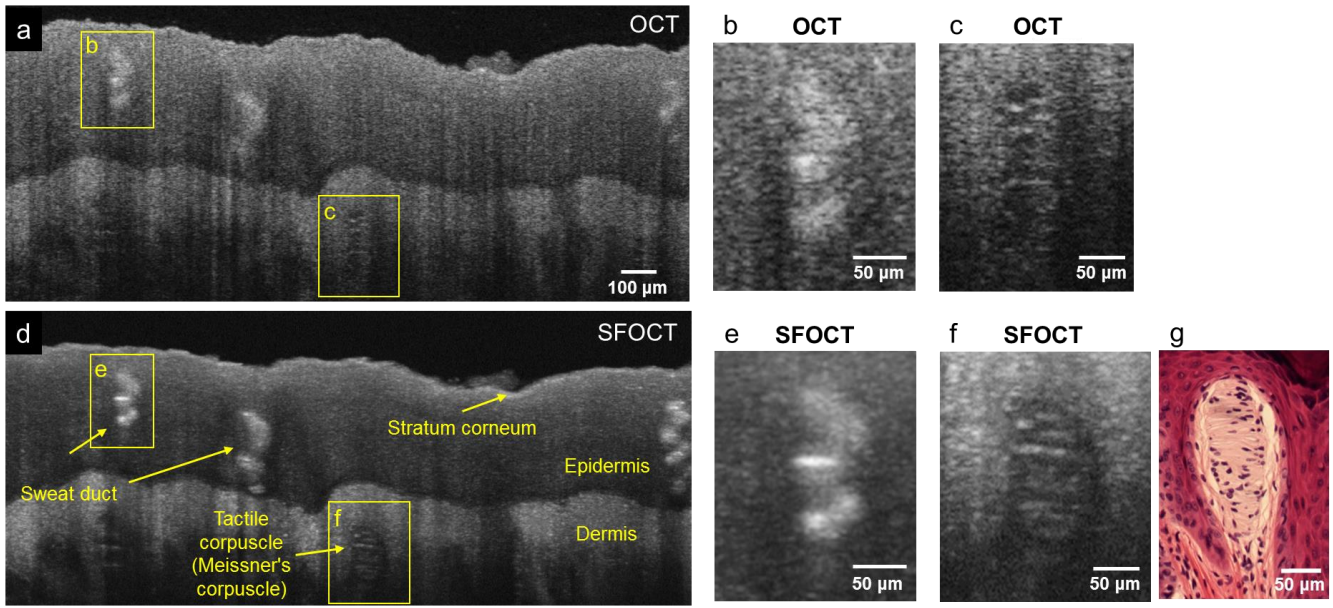

**Figure 5 | SFOCT imaging of intact human fingertip skin reveals fine structures such as the sweat ducts and the tactile corpuscle. a,** OCT B-scan of a fingertip. **b,** Close-up view on the sweat duct marked in **a**. **c,** Close-up view on the tactile corpuscle marked in **a**. **d,** SFOCT scan of a fingertip. **e,** Close-up view on the sweat duct marked in **d**. **f,** Close-up view on the tactile corpuscle marked in **d**. **g,** Microscope image of H&E stained tactile corpuscle [36].

# Supplementary Information

## Speckle-Free Coherence Tomography of Turbid Media

**Authors:** Orly Liba[1,2,3,4], Matthew D. Lew[1], Elliott D. SoRelle[1,3,4,5], Rebecca Dutta[1,3,4], Debasish Sen[1,3,4], Darius M. Moshfeghi[6], Steven Chu[4,5,7], Adam de la Zerda[1,2,3,4,5]*

**Affiliations:**

[1]Department of Structural Biology, Stanford University, Stanford, CA 94305.

[2]Electrical Engineering, Stanford University, Stanford, CA 94305.

[3]Molecular Imaging Program at Stanford, Stanford, CA 94305.

[4]The Bio-X Program, Stanford, CA 94305.

[5]Biophysics Program at Stanford, Stanford, CA 94305.

[6]Byers Eye Institute, Dept. of Ophthalmology, Stanford University School of Medicine, Palo Alto, CA 94303.

[7]Departments of Physics and Molecular & Cellular Physiology, Stanford University, Stanford, CA 94305.

*Correspondence to:  adlz@stanford.edu

**This document includes:**

Materials and Methods

Figures S1-S17

Tables S1-S4

Reference list for this document

**Materials and Methods**

S1. Simulation of speckle reduction with local-random phase variations

A Monte-Carlo simulation was implemented to simulate the reduction of speckle noise in turbid media using local-random phase variations. Eq. S1 depicts an approximation of the signal intensity originating from a voxel inside a turbid sample.

$$I = \frac{1}{M}\sum_{m=1}^{M}\left|\sum_{n=1}^{N}a_n e^{i\varphi_n}e^{i\theta_{n,m}}\right| \qquad (Eq.\ S1)$$

$I$ is the pixel value after averaging $M$ scans obtained at different times and with different local phase shifts within the illuminating beam. $N$ is the number of scatterers inside a voxel, and for each scatter $n$ within that voxel, $a_n$ is its scattering amplitude (proportional to its amplitude reflection coefficient) and $\varphi_n$ is the phase delay due to its axial location. $\theta_{n,m}$ is the local phase shift of the illumination beam, which changes in time in SFOCT, at the location of scatterer $n$.

Fig. S1 shows that increasing the number of averages narrows the pixel value distribution, thereby, reducing speckle noise. In the case of a random number of particles in each voxel, the pixel intensity variation persists with increased averaging owing to the actual variation of scatterers within each voxel. This underlying variation is why the distribution does not continue to narrow when increasing the number of averages from 100 to 1000.

The Monte-Carlo simulation considers 10,000 voxels. The number of particles in each voxel is either 30 or randomly chosen from a Poisson distribution with λ=30. The scattering amplitude is uniform across all particles, and equal to 1. The phase contributed by each particle, $\varphi_n$, is a random variable with a uniform distribution between 0 and $2\pi$. In OCT $\theta_{n,m}$ is constantly equal to zero, in SFOCT it is a random variable with a uniform distribution between 0 and $2\pi$.

S2. Experimental setup

SFOCT was implemented by modifying two existing OCT systems: the Ganymede HR (Thorlabs) and a clinical retinal imaging device (iFusion , Optovue). Both are SDOCT systems. All SFOCT images except for the human retina images were acquired using the Ganymede HR (HR-OCT).

The implementation of SFOCT on the Ganymede HR appears in Fig. 1. The light source of the Ganymede HR is a super luminescent diode (SLD) with a center wavelength of 900 nm and a 200 nm full bandwidth (λ=800-1000 nm), which provides 2.1 μm axial resolution in water. The spectrometer acquires 2048 samples for each A-scan at a measured rate of 20.7 kHz. All image reconstruction and analysis was made with Matlab (Mathworks) on the raw data from the spectrometer. The first lens of the imaging system (LSM03-BB, Thorlabs) provides a lateral resolution of 8 μm (full-width-half-max, FWHM) and depth of field (DOF) of 143 μm in water. In the Ganymede HR, the diffuser was placed at the original focal plane of the OCT probe, and a new focal plane was projected by a 4f imaging system [1]. The 4f configuration was implemented using two similar lenses (LSM02-BB, Thorlabs) that provide a lateral resolution of 4.2 μm (FWHM) and DOF of 32 μm in water. Due to the extension of the sample arm and the addition of 2 lenses and the diffuser, the reference arm was extended by approximately 10 cm and dispersion compensation elements were added (2x LSM02DC, Thorlabs). The reference arm was extended by placing metal rods between the OCT probe and the reference mirror. OCT images were obtained with the SFOCT apparatus without the diffuser. In this configuration, light propagates through the 4f imaging system and the extended reference arm. OCT images obtained this way are of similar quality to the OCT images obtained with the original probe. The only difference from the original probe is a 9% loss of power on the sample (Fig. S4a) which reduces SNR but does not change the properties of speckle. The diffuser was placed in the focal plane of the first lens and held within a custom motorized mount with XYZ translation (based on CXYZ1, Thorlabs). The diffusers were moved by a motor (Z812, Thorlabs), back and forth along one axis, and controlled through computer software (Thorlabs APT, Thorlabs). The movement of the diffuser was always perpendicular to the direction of the B-scan. The diffuser was translated back and forth at 0.3 mm/s over a range of 6.5 mm and an acceleration of 1.5 mm/s/s. Change in the direction of the diffuser occurred during the scan only when acquiring large volumes. Such volumes were acquired three times and those were averaged to obtain a volume in which the effect of the moving diffuser was noticed throughout the volume.

The implementation of SFOCT on the iFusion appears in Fig. S2. The iFusion is based on the iVue SDOCT. The scan beam center wavelength is λ=840±10 nm and provides an axial and lateral resolution of 5 μm and 15 μm in the retina, respectively. Each frame is composed of 1024 A-scans that are acquired at 26 kHz. The images in this study (Fig. S16) were acquired in Retina Cross Line Mode with 44 B-scan averages and a software parameter set to include all frames in the average. The diffuser was placed in the conjugate image plane (Fig. S2b). A dichroic mirror, which is used for obtaining fundus images, was removed to make room for the diffuser. The diffuser was held within a thin fixed mount (LH-1T, Newport) which was attached to a motorized translation stage. The stage and the diffuser were moved along one axis, perpendicular to the direction of the B-scan, and controlled through computer software (Thorlabs APT, Thorlabs). The diffuser was translated at a speed of 1.5 mm/s over a range of 6 mm. All processing was done internally by the iFusion computer and software.

The diffusers used for all experiments are ground glass diffusers with anti-reflective coating on one side (DG10-1500-B and DG10-2000-B, Thorlabs). The 1500 and 2000 grit diffusers are 2 and 1 mm thick, respectively. The 3 μm lapped diffuser was created by further lapping a commercial 1500 grit diffuser with 3 μm aluminum oxide grit (Universal Photonics) for 15 minutes. The profile and height statistics of the diffusers appear in Fig. 1c-d and Fig. S3.

Accurately positioning the diffusive plane of the diffuser at the waist of the Gaussian beam of the OCT was crucial for obtaining high quality images. Deviations from the ideal diffuser axial location resulted in power and resolution losses and impeded the speckle reduction effect. We manually positioned the diffuser at this ideal location by changing the location of the diffuser along the optical axis until an optimal image was obtained. In the retinal OCT system, the optimal placement of the diffuser along the optical axis was found by a combination of two indications. First, the person being examined adjusted the location of the diffuser until an optical test target (inherent to the iFusion) appeared in sharp focus to the examinee. Next, the person acquiring the images adjusted the location of the diffuser to obtain an optimal signal to noise ratio (SNR) of the image of the retina shown on the computer screen.

S3. Design considerations and characterization of ground glass diffusers

To achieve maximal phase decorrelation the random phases added by the diffuser, $\theta_{n,m}$ (Eq. S1), should be evenly distributed between 0 to $2\pi$ at the OCT focal plane. In order to obtain this phase shift using a diffuser made of glass with a refractive index of 1.5 (NBK-7) and light sources with a center wavelength of 900 nm, the total thickness variation of the diffuser should span at least 1.8 μm. However, deflection of light by the diffuser, which is more probable in a ground glass diffuser with a large thickness variation, reduces the OCT signal and should be minimized. In our implementation, we used three types of diffusers and characterized their thickness and roughness (Fig. 1c-d, Fig. S3) with a 3D optical profiler. The roughest diffuser is a commercial 1500 grit diffuser with anti-reflective coating. The finest diffuser was made by further grinding (lapping) the 1500 diffuser with 3 µm particles (3 µm lapped diffuser). We also used a 2000 grit diffuser, which has a roughness between the previously mentioned diffusers. Each of the three diffusers tested had a small effect on the optical power on the sample (Fig. S4a, Table S2), the OCT signal (Fig. S4b, Table S3), and the lateral resolution (Fig. S5, Table S4). While the roughest diffuser (1500 grit) reduced the OCT signal and the lateral resolution more than the other diffusers, it achieved the best qualitative performance in terms of speckle removal and appearance of fine anatomic detail in tissue.

The profiles of the diffusers were measured with a non-contact 3D optical profiler (S neox, Sensofar). The profiles were obtained with a 50x magnification objective lens (Nikon, NA 0.55 DI) in an interferometric scan mode. Post processing was performed with the SensoSCAN program (Sensofar) and included depth slope correction and calculation of the depth histogram. Additionally, the profile of the 1500 grit diffuser required restoration (interpolation) due to regions from which light was not collected.

S4. Post processing

The post-processing methods in this section were applied only for the HR-OCT system (Ganymede HR, Thorlabs). Post processing was done with Matlab (2015a and 2014b, Mathworks). The raw spectrum of each A-scan acquired with OCT and SFOCT was processed in a similar way to create images in the spatial domain. Reconstruction was performed by subtracting the spectrum of the source, as measured by the OCT, and multiplying by a phase matrix that is equivalent to applying a Fourier transform [2]. To reduce spectrum-derived artifacts, the spectrum was multiplied by a Hann window with 2048 points. Prior to reconstruction, dispersion compensation was performed by finding the coefficient of the quadratic phase term iteratively by minimizing the absolute difference between the reconstructed images of two distinct spectral windows [3]. Dispersion compensation was done separately for each experiment. To obtain the final OCT/SFOCT image, the magnitudes of multiple reconstructed B-scans were averaged on a linear scale.

In order to minimize movement artifacts, frames that were notably different from most of the frames in the scan were excluded from averaging. This exclusion process was done only for the mouse retina and mouse cornea images, in which there was significant movement due to breathing. Frame similarity was determined by measuring the correlation of each frame to the average of all the frames. The threshold for excluding frames was determined manually for each scan.

The number of averages for each image appears in Table S1. The averaged image is displayed on a logarithmic scale with image-adaptive brightness scaling unless otherwise stated. Dark pixels correspond to low scattering from the sample while bright pixels correspond to high intensity of scattering.

S5. Phantom Preparation

The PDMS-TiO$_2$ phantom (Fig. 1f-i) was fabricated by spin-coating layers of PDMS (Sylgard 184 Silicone Elastomer, Dow/Corning) comprising TiO$_2$ powder particles (TiO$_2$ anatase, 232022, Sigma-Aldrich) with an average size of 130±70 nm.

Agarose phantoms embedded with various scattering agents were created using a stock solution of 1% agarose (J.T. Baker) in water. Two different scattering agents were used: $TiO_2$ anatase nanopowder with 21 nm primary particle size (Sigma Aldrich) and large gold nanorods (LGNRs) with peak absorption at 745 nm and size of approximately 90 by 35 nm [4]. LGNRs were used because their scattering to absorption ratio is higher compared to conventional gold nanorods [5]. The agarose-LGNR phantom without beads (Fig. S8) consisted of $10^{11}$ LGNRs/mL. The agarose-LGNR phantom with beads (Fig. 2 e-k, S10a) consisted of $2x10^{11}$ LGNRs/mL and polystyrene beads (Streptavidin Polystyrene Particles, average diameter 3.05 µm, 0.5% w/v, Spherotech) at a final concentration of $2.38x10^8$ beads/mL (0.04 beads/voxel). The agarose-$TiO_2$ phantom (Fig. S10 b-d) was fabricated by dispersing 0.009 grams of nanopowder in 1 mL ultrapure water. The solution was sonicated, however, the clumps persisted. For the three phantoms described above, the scattering agents and polystyrene beads were slowly added to 5 mL of uncured agarose at 60 °C with continuous stirring. The final solution was allowed to stir for one minute before being poured into 5mL plastic petri dishes. The phantoms were allowed to cure for at least two hours before imaging.

S6. Optical power and signal intensity

The optical power on the sample and the OCT signal were measured for OCT and SFOCT with the three different diffusers on the HR-OCT (Fig. S4). The optical power was measured by placing a power meter (PM122D, Thorlabs) with a germanium sensor (S122C, Thorlabs) and aperture 9.5 mm at the focal plane of the scan lens while scanning at a single point at the center of the field of view. The measurement was calibrated for the center wavelength of the source, 900 nm. At least 100 consecutive measurements were acquired with the power meter for a time period of ~60 seconds. The OCT measurement refers to the original probe without any additional components. The measurement named "no diffuser" refers to the SFOCT system, which adds two lenses to the original probe, without a diffuser. The signal intensity was measured on images of a PDMS+$TiO_2$ phantom with 100 B-scan averages. The regions (500 µm long and 100 µm deep) selected for the measurements were all chosen at the same depth in the phantom, location relative to the focal plane, and position on the screen, to eliminate the effect of absorption, focusing and signal roll-off. The values are on a linear scale and in arbitrary units. The relatively high standard deviation in the signal intensity is due to aggregations of $TiO_2$, distance from the focal plane and absorbance inside of the region selected for this measurement. Note that the decrease in signal intensity due to the diffusers is larger than the decrease in power on the sample, because the signal is created by light that is travelling twice through the diffuser and because some light that was measured by the power meter at the sample will be rejected via the confocal detection of our OCT system.

S7. Measurement of lateral resolution

The lateral resolution of OCT and SFOCT was evaluated using a 1951 USAF glass slide resolution target (Edmund). The target was placed at the focal plane and scanned in 3D mode three times, each with 18 B-scan averages, to obtain a total of 54 B-scan averages. The samples were scanned with 3 µm spacing in both lateral directions over a 1x1 mm square. The reconstructed volumes were averaged on a linear scale to create a single volume for each condition. Next, 200 rows that include the surface of the resolution chart were averaged along the z axis to obtain an *en face* projection of the volume. The images are displayed on a logarithmic scale (Fig. S5).

For each type of scan, the smallest resolvable group was determined visually in both horizontal and vertical directions. The inverse of the line-pairs-per-mm of the smallest resolvable group was used to calculate the effective full-width-half-max (FWHM) of the beam. The PSF size increase is calculated as 1-[(SFOCT average lines-pairs-per-mm)/(OCT average lines-pairs-per-mm)]. Note that the FWHM of the OCT, as defined by the scanning lens, is 11.2 µm in air. The effective resolution is better because we are determining the visibility of line separations visually on logarithmic scale images.

S8. Measurement of gap in phantom

Images of the PDMS-$TiO_2$ phantom acquired with a bright field microscope (10x, NA=0.25), OCT and SFOCT were registered and segmented to detect the gap in the phantom (Fig. 1f-i, S6, S7). Identical cross-section planes (*en face*) were chosen from the 3D scans acquired with OCT and SFOCT. Because these scans were taken consecutively, only minor translation adjustments were required in order to register the cross-sections. The microscope image was scaled using the known pixel size ratio between the OCT and microscope, and this image was registered to the OCT and SFOCT scans manually with translation and rotation (Fig. S6h). The segmentation threshold of the OCT and SFOCT scans was chosen carefully so that a maximal amount of pixels will be considered inside the gap, but without large holes inside the phantom. Small holes were filled with the morphological close operator with a structure element of size 2 (imclose), which is equivalent to 4 µm. The threshold was chosen as the lowest value in which the phantom had no holes inside it. After finding the segmentation threshold, the original image was segmented, without morphological closing (Fig. 6e-g). In order to exclude holes in the measurement of the gap while retaining its shape and size, holes were filled using the imfill function. Segmentation of the microscope image was performed using Otsu's method (graythresh) function. The gap was measured by counting the number of pixels belonging to the gap across the x axis of the phantom (Fig. S7).

S9. Statistical analysis

OCT speckle noise follows a Rayleigh distribution (Eq. S2). SFOCT reduces speckle and shifts the pixel value statistics from a Rayleigh distribution toward the expected distribution of scatterers in a sample: a Poisson distribution [6], in which each scattering event contributes a signal value that is equal to the backscattering of a single LGNR (Eq. S3). Eq. S3 was used as a simplified model to approximate a Poisson curve to the pixel value distribution; it does not present a complete model or proof for the pixel value distribution. For simplicity, we assume that every individual LGNR produces equal backscattering, and we also neglect intensity changes due to scattering, absorption, and distance from the optical axis and the focal plane. However, these effects do contribute to the pixel value distribution, in addition to the spatial distribution of LGNRs.

The expressions for the two distributions are:

$$p_{rayleigh}(I) = \frac{\pi I}{2\langle I \rangle^2} \exp\left[-\frac{\pi I^2}{4\langle I \rangle^2}\right] \tag{Eq. S2}$$

$$p_{poisson}(k) = \frac{\lambda_p^k e^{-\lambda_p}}{k!}, \quad p_{poisson}(I) = I_p \times p_{poisson}(k) \tag{Eq. S3}$$

in which $I$ is the pixel value and $\langle I \rangle$ the average pixel intensity. $k$ is the number of particles in a voxel, each of which we assume contributes equally to the OCT signal and $\lambda_p$ is the average number of particles in a voxel. In order to obtain $\lambda_p$ we fit the histogram of the pixel values to a Poisson distribution. From that we derive that the contribution of each equivalent particle to the OCT signal is $I_p = \langle I \rangle / \lambda_p$. Fitting the Poisson curve to the distribution was done by iterating over possible values of $I_p$, calculating $\lambda_p = \langle I \rangle / I_p$ and choosing the Poisson curve, $p_{poisson}(I)$, with the minimal mean squared error (MSE) compared to the measured pixel value distribution. This method yielded a smooth curve of the MSE versus $I_p$ and a global minimum.

Statistical analysis of the pixel values with OCT and SFOCT (Fig. 2a-b, S9a) was performed by analyzing images of an agarose-LGNR phantom (Fig. S8). The 2D ROI (1 mm long and 200 µm deep) was chosen at a similar depth inside the phantom, at a narrow region around the image plane. The statistical analysis was performed on the pixel values on a linear scale.

S10. Imaging of live mouse pinna, retina and cornea

All animal experiments were performed in compliance with IACUC guidelines and with the Stanford University Animal Studies Committee's Guidelines for the Care and Use of Research Animals.

Nu/nu mice (Charles River Labs) were anesthetized by inhalation of 2.5% isoflurane in $O_2$ (v/v). Once adequately anesthetized, the right ear pinna was immobilized using double-sided tape. We optimized light transmission to the sample by applying ultrasound (US) gel to the mouse skin and covering the gel with a 2 mm thick one sided anti-reflective coated glass (650-1050 nm), with the coating at the air-glass interface.

For retinal and corneal imaging studies, nu/nu mice were anesthetized with intraperitoneal injections of 80 mg/kg ketamine (Vedco Inc.) and 10 mg/kg xylazine (Lloyd Inc.). Once adequately anesthetized, the mice were mounted on to a platform and secured with stereotactic devices on both sides of the skull. With the mouse secured, the stage was tilted to orient the mouse's eye upward, with the top of the cornea being approximately parallel to the table. Pupillary dilation was achieved by applying 1 drop each of 1% tropicamide (Bausch & Lomb), and 2.5% phenylephrine hydrochloride (Paragon BioTeck) to the eyes, for 2 min each. 2.5% hypromellose solution (Gonak™, Akorn Inc.) was then placed over each eye as a contact solution. In this position, a plastic 'O'-ring was carefully put surrounding the eye using vacuum grease (Dow Corning), to avoid contact with the cornea. Finally, 0.17 mm thick glass cover slips were gently placed over the 'O' ring to complete a fluid-tight chamber over the dilated eye. Anesthesia was continually maintained using a nose-cone delivering 1.5-2% isoflurane in $O_2$.

S11. Imaging of human fingertip

The fingertips of healthy volunteers were imaged with the Ganymede HR. The volunteers' arms were immobilized, and the subject's finger was pressed onto the bottom of a fixed glass window with anti-reflective coating at the air-glass interface. As in the setup for imaging the mouse pinna, we optimized light transmission to the sample by applying US gel to the fingertip skin. To minimize movement, SFOCT scans were acquired in 2D or 3D modes with the diffuser, and then the diffuser was quickly removed to acquire the corresponding conventional OCT images.

S12. Imaging of human retina

A healthy human volunteer (one of the authors) was scanned with the retinal system (iFusion, Optovue). Stanford Office of Human Subjects Research determined that this did not warrant for an institutional review board. The diffuser was placed in the image plane for SFOCT imaging and removed for OCT imaging. Images were acquired quickly one after the other to reduce movement between scans.

S13. Alternative methods of speckle reduction

SFOCT was compared to common alternative speckle reduction techniques. Spatial compounding, one of the most common methods for reducing speckle, was implemented in two ways. First, by performing 3D Gaussian smoothing on a volume acquired with conventional OCT (Fig. S11 c-d). Smoothing was performed on the linear-scale image (after resampling) using Matlab's smooth3 function with a Gaussian kernel in a square window with a size of 11 pixels. The standard deviation of the Gaussian was 0.95 and 1.25. The second method attempted to reduce speckle by averaging adjacent frames along the y axis (Fig. S11 e-h). Averaging was performed on the linear-scale images over 4, 7, 9, and 13 frames, respectively (spanning 12, 24, 32, and 48 µm). These images show that although spatial compounding reduces speckle noise, it does not reveal structure as well as SFOCT.

Polarization diversity is expected to provide minimal speckle noise reduction due to the limited number of possible uncorrelated speckle patterns[7,8]. Frequency compounding requires reconstructing the OCT image from several sub-bands. Compounding a large number of non-overlapping bands would reduce the axial resolution considerably. Therefore, these two methods were not examined. Angular compounding often produces favorable results however compounding many angles reduces resolution and DOF.

Digital removal of speckle noise is also widely used. We compared SFOCT images to OCT images that were post-processed with one of three digital methods with several different parameters. The filters were applied on a logarithmic scale OCT image [9]. Adaptive Wiener filtering was implemented with the Matlab function wiener2, using neighborhoods of sizes 5x5 and 7x7 pixels (Fig. S12 c-d). Hybrid median filter (HMF) was implemented [10] using neighborhoods of sizes 5x5, 7x7 and 9x9 pixels (Fig. S13 a-c). Symmetric nearest-neighbor filter (SNN) was implemented [11] using neighborhoods of sizes 5x5, 7x7 and 9x9 pixels. These examples show that filters of a small size do not reduce speckle noise sufficiently, whereas, larger filters create artifacts and compromise some of the information in the image.

S14. Supplementary discussion

As is true for most compounding methods, SFOCT requires averaging of several OCT images and therefore extends the time of image acquisition. However, the benefit of detecting small anatomical features for diagnostic imaging is worth the cost of increased acquisition time. In this manuscript, most of the images we presented are a result of compounded frames. With a fast moving diffuser and a modified acquisition pattern, it is possible to average A-scans (Fig. S17). In this case the diffuser should move more than one wavelength per scan, which is about 30 mm per second for an OCT which scans at 30 kHz. Note that the diffuser should not move too fast in order to avoid washout of the interference fringes.

One may intuit that the addition of the diffuser would degrade the lateral and axial resolution and significantly reduce the signal to noise ratio. Indeed, the addition of random local phase shifts introduces aberrations to the illuminating wavefront, which will inevitably broaden the waist of the focused beam and reduce the lateral resolution. However, as described below, the measurements presented in this study show that the compromise in resolution in SFOCT is insignificant. Moreover, the resolution of SFOCT images of turbid media is consistently superior to that of OCT images owing to the removal of speckle noise.

The axial resolution of the system is preserved owing to the averaging of the thickness variations in time. This averaging is why features as small as 2 μm are resolved along the vertical direction (such as the dark line in Fig. 3f).

The lateral resolution of SFOCT was measured using a resolution test target. The results show an increase of 35.3 % and 6.2 % in the size of the point spread function with the 1500 grit and 2000 grit diffusers, respectively (Fig. S5, Table S4). We believe that the difference between the two diffusers was caused by the increased roughness of the 1500 grit diffuser, which resulted in deflection of light and local loss of signal. The local signal variations were not completely removed by the 54 averages used to create these images. While the lateral resolution of SFOCT was degraded compared to OCT when measured on a glass test chart, the effective resolution of SFOCT was superior when imaging turbid samples, which are ordinarily dominated by speckle. We have measured a 250% increase in the effective resolution measured by SFOCT compared to conventional OCT (Fig. 1j). The achievable increase in visibility owing to SFOCT may be much higher in features with lower contrast, which are usually completely hidden by speckle noise in conventional OCT.

The optical power and signal intensity were measured for all three diffusers and compared to conventional OCT (Fig. S4, Table S2, S3). In the HR-OCT system, the power on the sample decreases by 9% due to the 4f imaging system and an additional reduction of 22%, 20%, and 25% due to the 3um lapped, 2000 grit and 1500 grit diffusers, respectively. The OCT and SFOCT signals were measured from a PDMS-TiO$_2$ phantom and yielded an average signal loss of 36%, 42%, and 50% with the three diffusers compared to OCT. While the current implementation of SFOCT suffers from a lower signal to noise ratio, the technique is still able to produce images with better visibility of fine features compared to OCT. Note that the noise in pixels within the tissue was much higher than that within pixels above the tissue (Fig. 2 a-b, for example), indicating that speckle noise was more significant than photon shot noise and thermal noise in our measurement (after several averages). In most applications, the reduction in signal intensity can be compensated by increasing the incident power on the sample.

In the retinal OCT system described herein, in addition to not being able to increase the power of the light source (due to practical limitations of the instrument), we did not have access to the raw scan data. Thus, we were unable to optimize the acquisition and post- processing algorithms in order to produce higher quality images with the iFusion. To circumvent these limitations, we created a finer grit diffuser by lapping a 1500 grit diffuser with 3 µm particles. The finer diffuser produced images with a higher signal to noise ratio but possibly at the expense of effective speckle noise reduction. Although speckle may still be present in the retina images, the retinal layers are better defined in SFOCT when compared to conventional OCT (Fig. S16).

# Figures

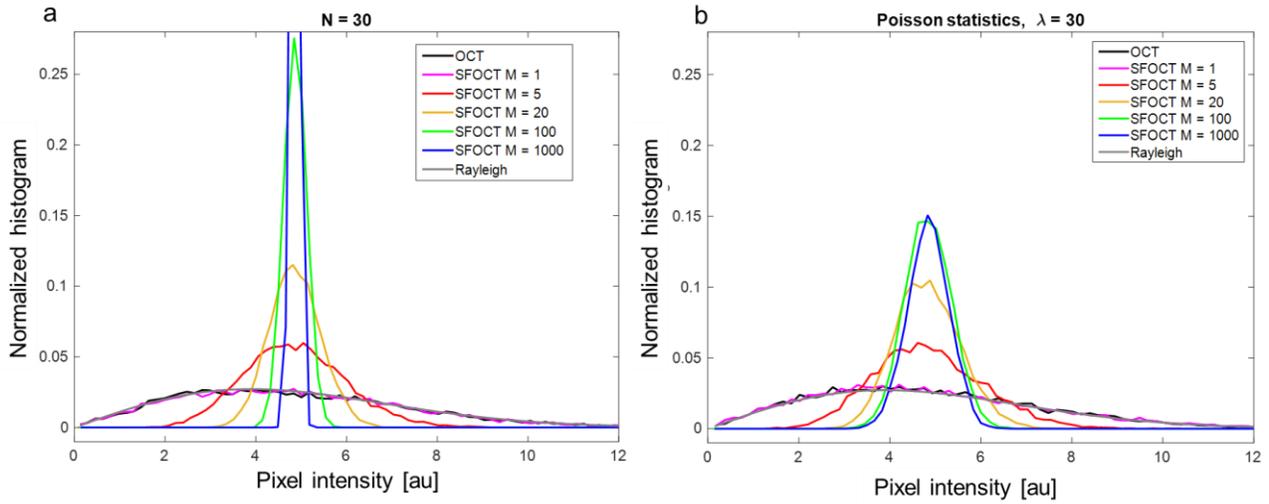

**Figure S1| Normalized histograms of the pixel values obtained with a Monte-Carlo simulation of Eq. S1 with 1, 5, 20, 100 and 1000 averages. a,** The results of the simulation with 30 particles in each voxel. **b,** The results of the simulation with the number of particles in each voxel randomly chosen from a Poisson distribution with λ=30. Due to the random distribution of particles, the variance of pixel values does not reduce to zero. This explains why the images of phantoms composed of particles dispersed in a volume do not appear completely uniform with SFOCT even with 100 averages.

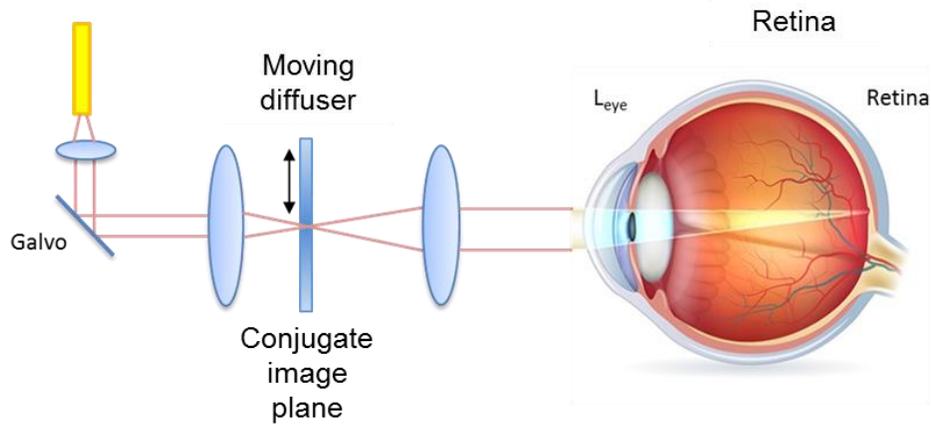

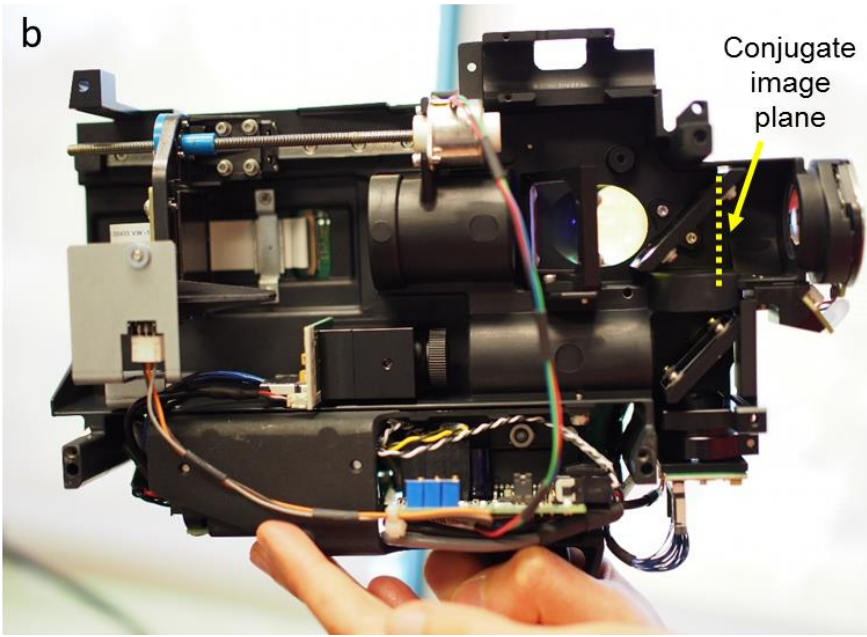
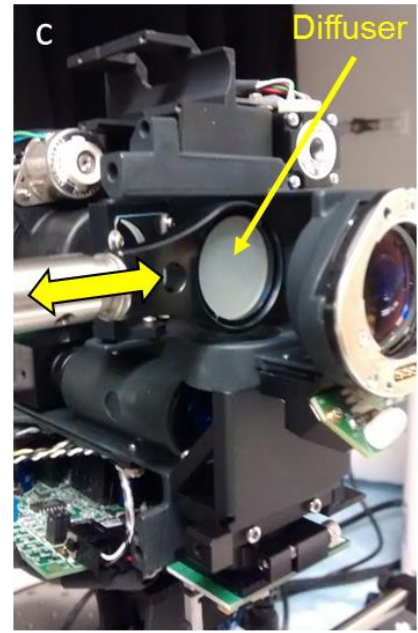

**Figure S2 | Schematic and implementation of SFOCT for a human ophthalmic imaging system. a,** A schematic showing the placement of the moving diffuser in the optical path of an ophthalmic OCT. **b,** The interior of the scan-head of the iFusion. The conjugate image plane is marked by the dashed yellow line. **c,** The scan-head with the diffuser placed in the conjugate image plane. The yellow arrow shows the direction of the motion of the diffuser.

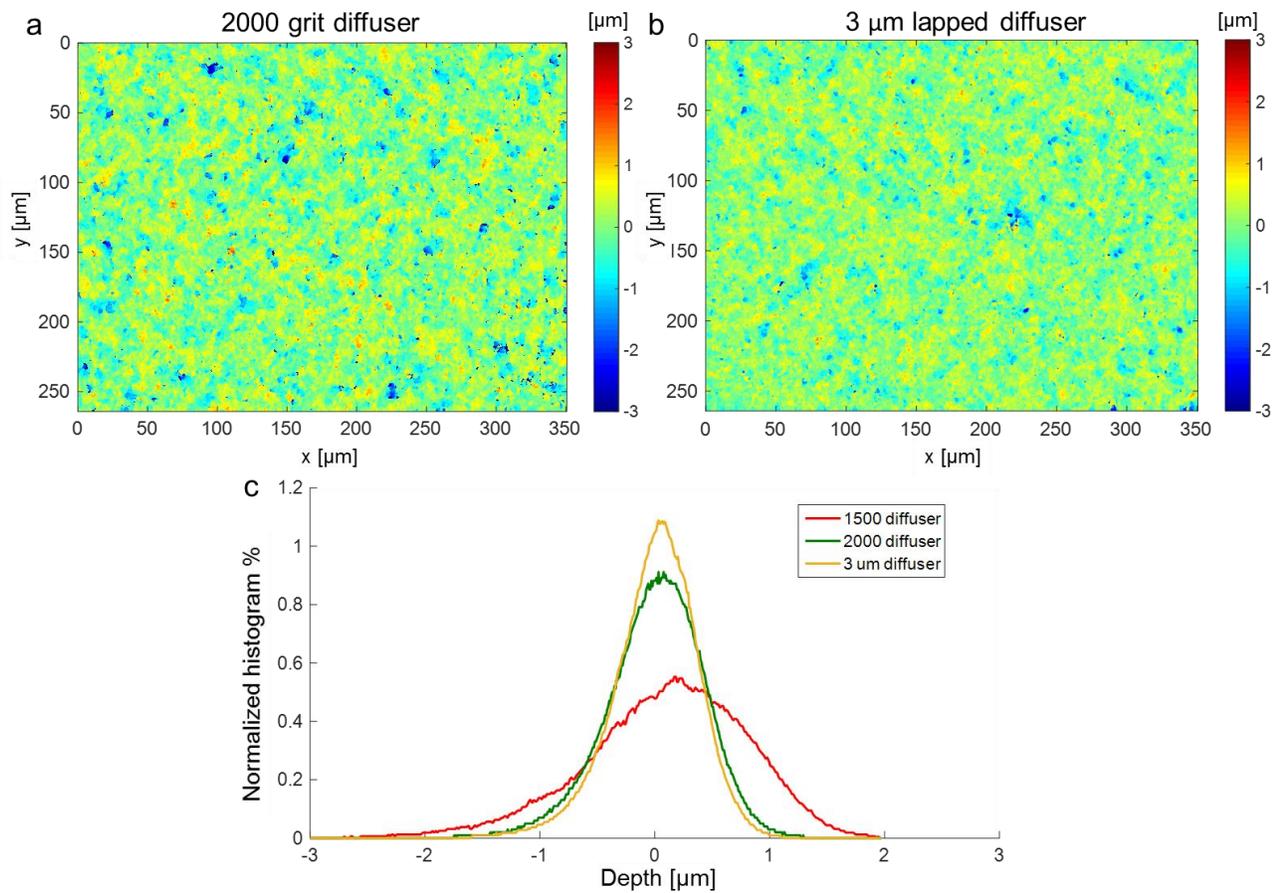

**Figure S3| Characterization of the thickness profile of the diffusers. a,** Depth profile of the 2000 grit diffuser. **b,** Depth profile of the 3 μm lapped diffuser. **c,** Depth histogram of all of the diffusers used in this study. The depth profile of the 1500 grit diffuser is shown in Fig. 1c.

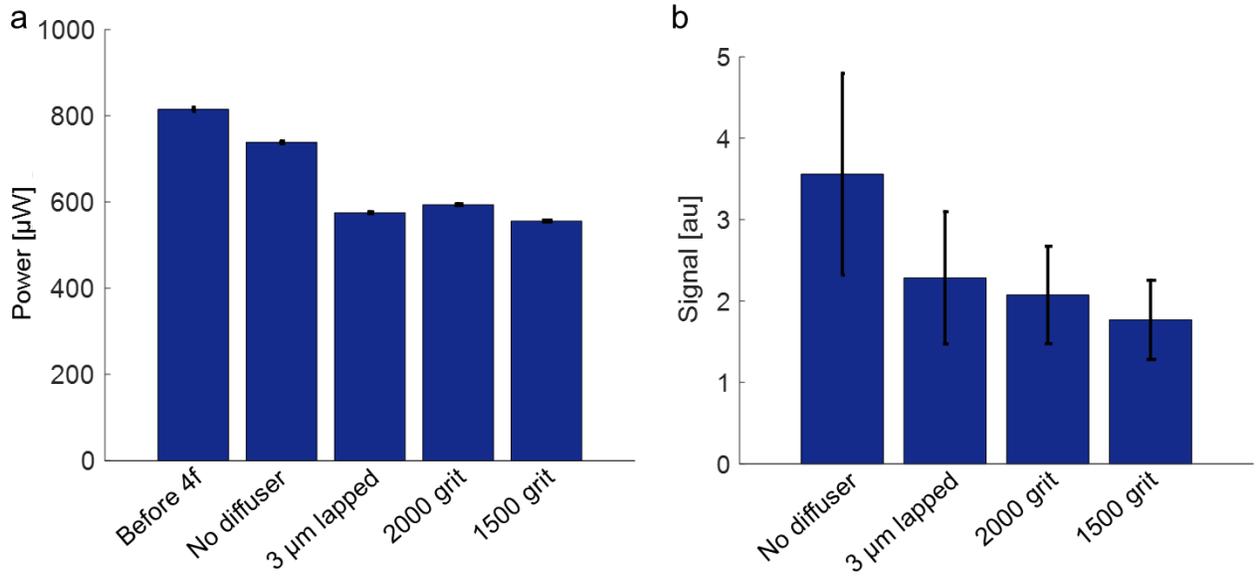

**Figure S4| Power and signal intensity of OCT and SFOCT with the three diffusers. a,** Power on sample for OCT and SFOCT. **b,** Signal intensity of OCT and SFOCT. Error bars are the standard deviation of the measurements.

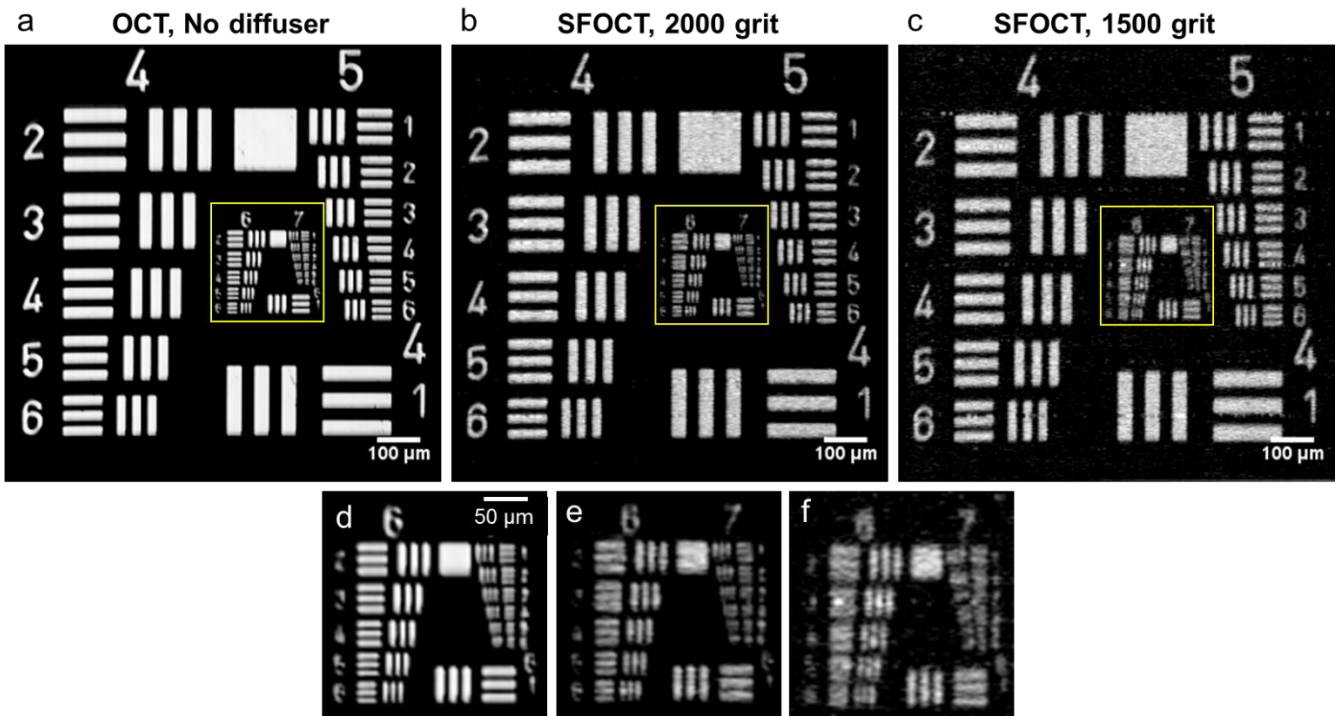

**Figure S5| Lateral resolution of OCT and SFOCT with the two diffusers, measured on a 1951 USAF resolution test target**. **a, d,** OCT, **b, e,** SFOCT with 2000 grit diffuser, and **c, f,** SFOCT with 1500 grit diffuser.

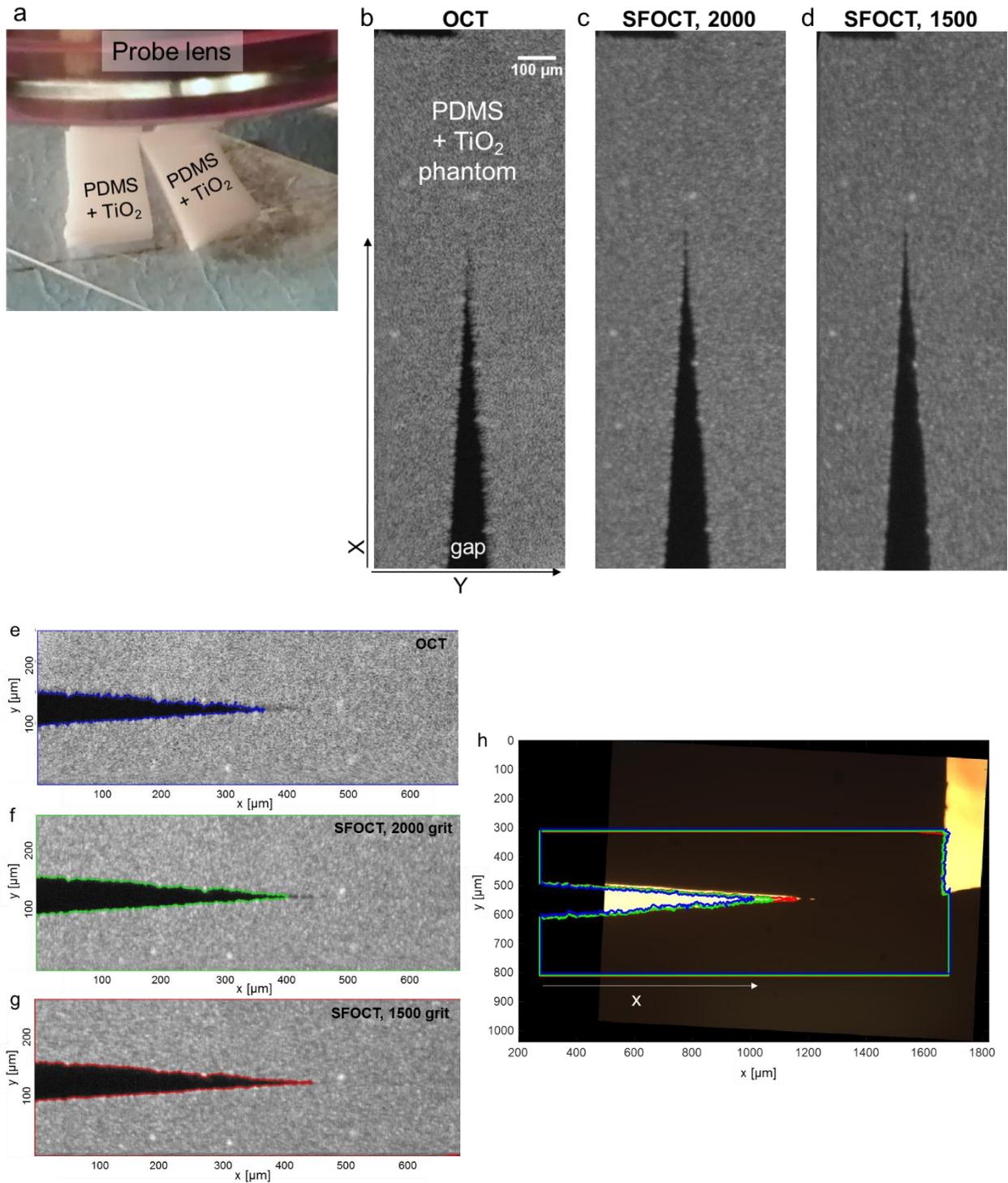

**Figure S6| Description of the measurements used to determine the effective resolution in a turbid sample with OCT and SFOCT**. **a,** A view of the phantom under the imaging system, showing the two parts of PDMS and the gap between them. **b,** OCT *en face* image inside of the phantom. **c,** Similar scan using SFOCT with 2000 grit diffuser. **d,** Similar scan using SFOCT with 1500 grit diffuser. **e-g,** OCT and SFOCT scans along with the lines which represent the segmentation boundary between the phantom and the gap. **h,** The segmentation boundaries drawn on top of a registered image of the phantom obtained with a bright-field microscope (10x, NA=0.25).

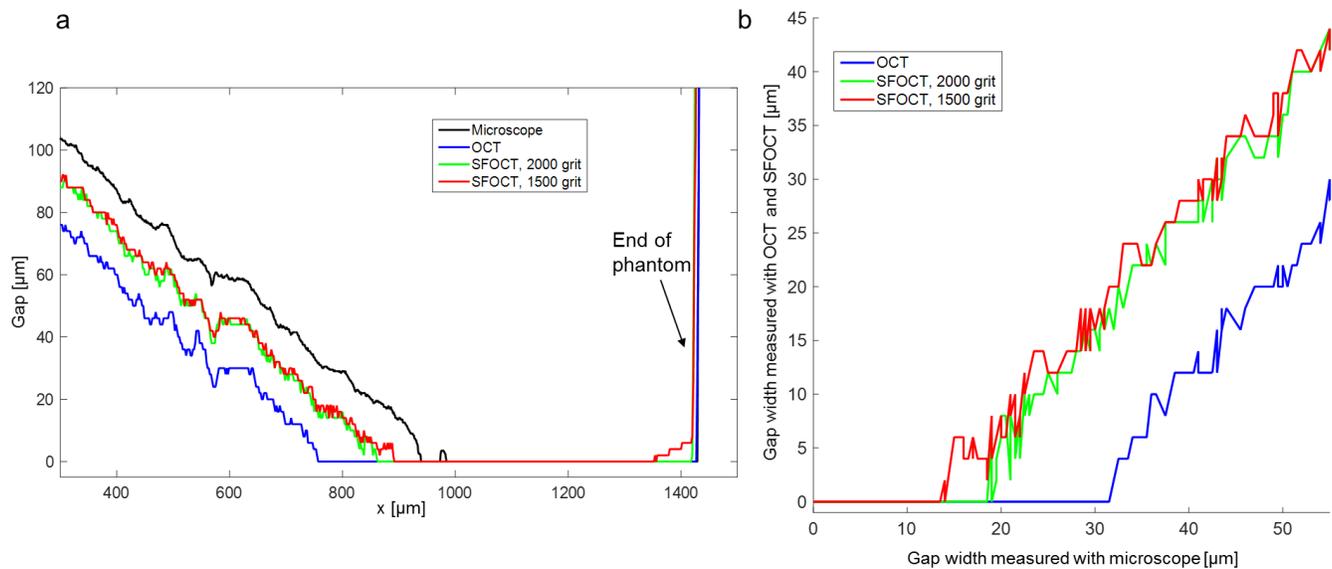

**Figure S7| The width of the gap in the turbid phantom, measured by OCT, SFOCT and a bright field microscope**. **a,** The size of the gap as a function of location (x) as measured from Fig. S6 h. **b,** The size of the gap in the OCT and SFOCT images, as shown in **a**, plotted as a function of the size of the gap in the microscope image. The difference in visibility between OCT and SFOCT is clearly shown. The effective resolution in SFOCT is 2.5-fold better than OCT.

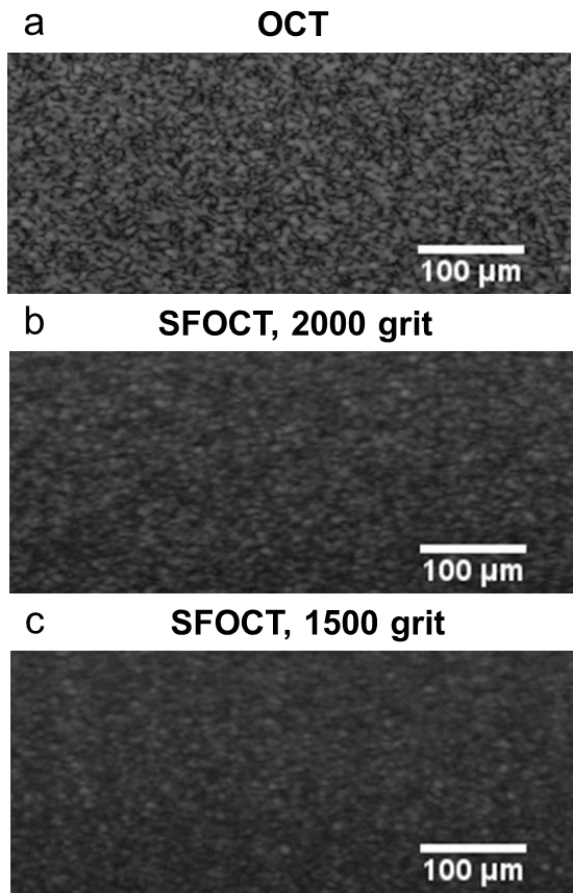

**Figure S8| OCT and SFOCT images of the agarose-LGNR phantom. a,** OCT, **b,** SFOCT with 2000 grit diffuser, **c,** SFOCT with 1500 grit diffuser. The OCT image shows a combination of speckle noise and the signal variation due to the random distribution of LGNRs in the phantom. The SFOCT images show only the latter. This claim is supported by the statistical analysis of pixel intensities.

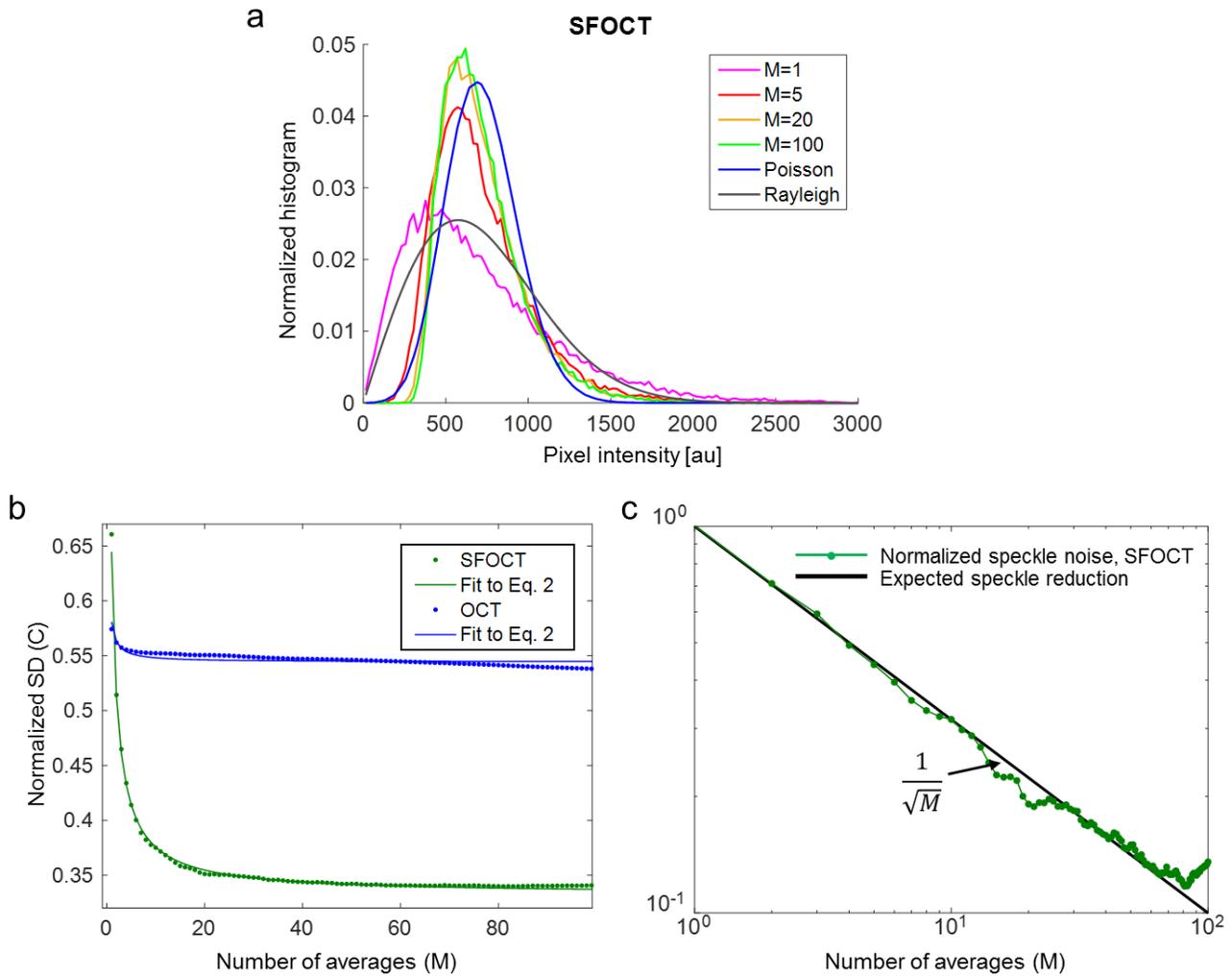

**Figure S9| Analysis of pixel values obtained with SFOCT with a 2000 grit diffuser, performed similarly to Fig. 2b-d. a,** Normalized histogram of pixel values showing the transition from speckle statistics (Rayleigh) towards a Poisson distribution as the number of averages, M, increases. The Poisson distribution expresses the probability of a given number of LGNRs to be present in a single voxel. **b,** Reduction in normalized SD versus the number of averages, M, for OCT and SFOCT. The reduction in the normalized SD is significantly larger in SFOCT versus OCT, and follows the theoretical dependence on M as expressed in Eq. 2. **c,** The reduction of normalized speckle, as defined by Eq. 3, follows $1/\sqrt{M}$, as expected.

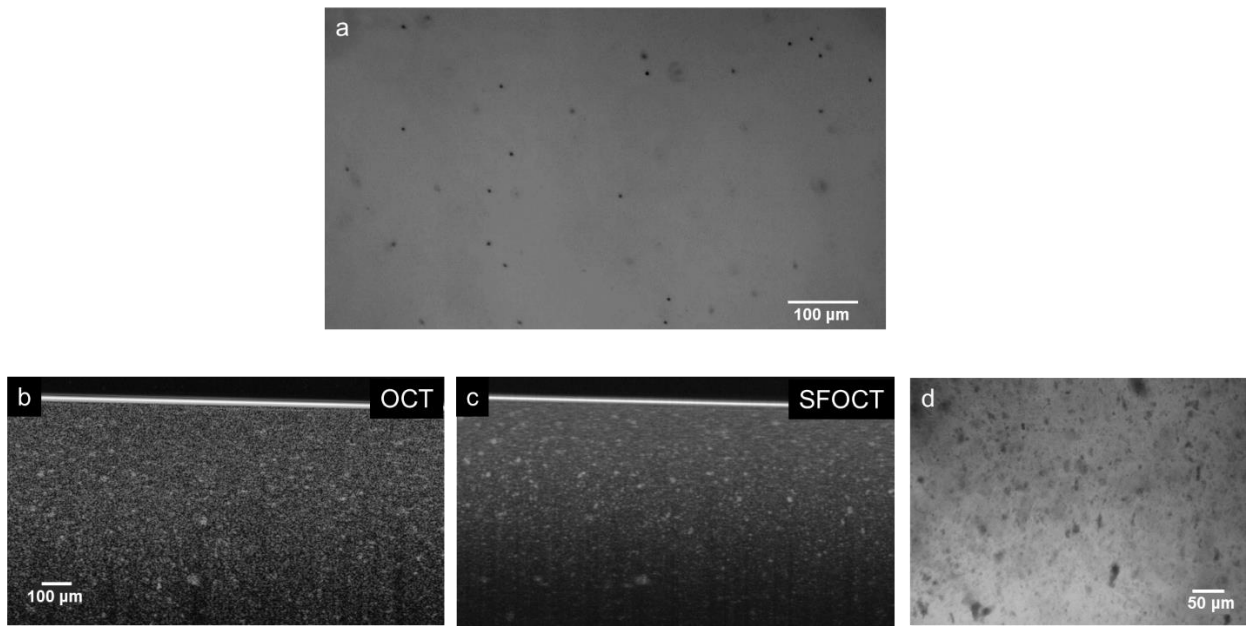

**Figure S10| Validation of the capability of SFOCT to produce images that better represent the true structure of the sample. a,** Intensity image (gray-scale) from a bright field microscope of a thin slice of an agarose-LGNR phantom with 3 µm diameter beads (10x, NA = 0.25). The image shows the beads sparsely dispersed in a uniform phantom. The LGNRs are too small to be resolved individually. **b, c,** OCT and SFOCT cross-sectional scans of an agarose phantom with $TiO_2$ nanopowder. The nanopowder does not disperse well in the phantom; rather, it forms large clumps. These clumps are hidden within the speckle noise in the OCT image, but are revealed in the SFOCT image. **d,** A bright field microscope (gray-scale) image of the phantom presented in **b, c**, showing the shape size and distribution of the clumps in the agarose base. The comparison of SFOCT images to microscope images shows that SFOCT provides a close representation of the true structure of the phantoms.

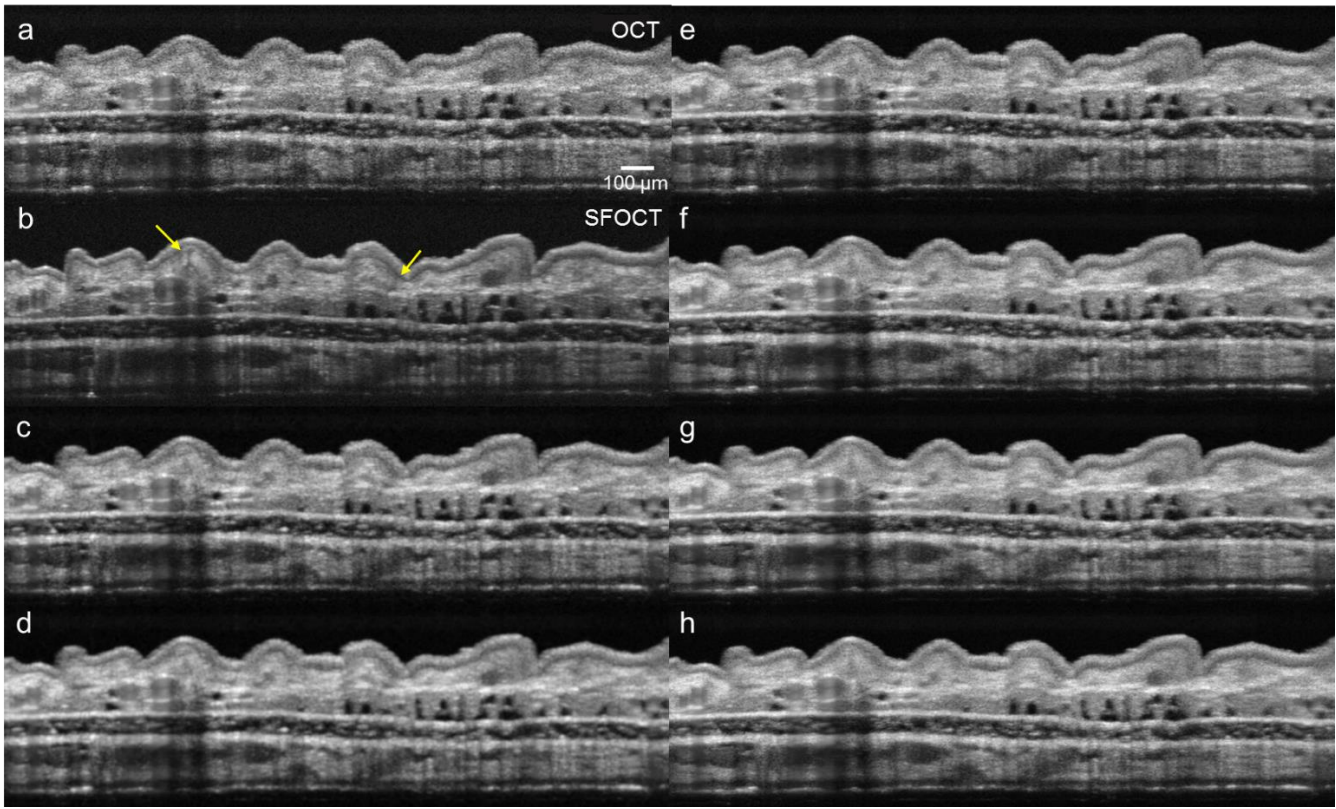

**Figure S11| Comparison of SFOCT to spatial compounding and 3D smoothing**. **a,** OCT scan sampled every 4 μm in both lateral directions, with 20 B-scan averages. The image was created by averaging 2 adjacent frames (to improve SNR, averaging is of linear-scale images) and later resampled to obtain a voxel size of 2 μm in all three directions. Averaging 2 scans, which span 4 μm, does not reduce speckle because the averaged scans are inside the PSF. **b,** SFOCT scan, with the 1500 grit diffuser, using the same acquisition parameters and post processing as **a**. **c, d,** Images resulting from three-dimensional smoothing of the OCT volume described in **a**. Smoothing is done on the linear-scale image after resampling using Matlab's smooth3 function with a Gaussian kernel in a square window with a size of 11 pixels. The standard deviation of the Gaussian is 0.95 in C and 1.25 in **d**. **e-h,** Averaging and processing of the OCT volume as described in **a**, of 4, 7, 9, and 13 frames, respectively (spanning 12, 24, 32, and 48 μm). These images show that although spatial compounding reduces speckle noise, it does not reveal structure (yellow arrows) as well as SFOCT.

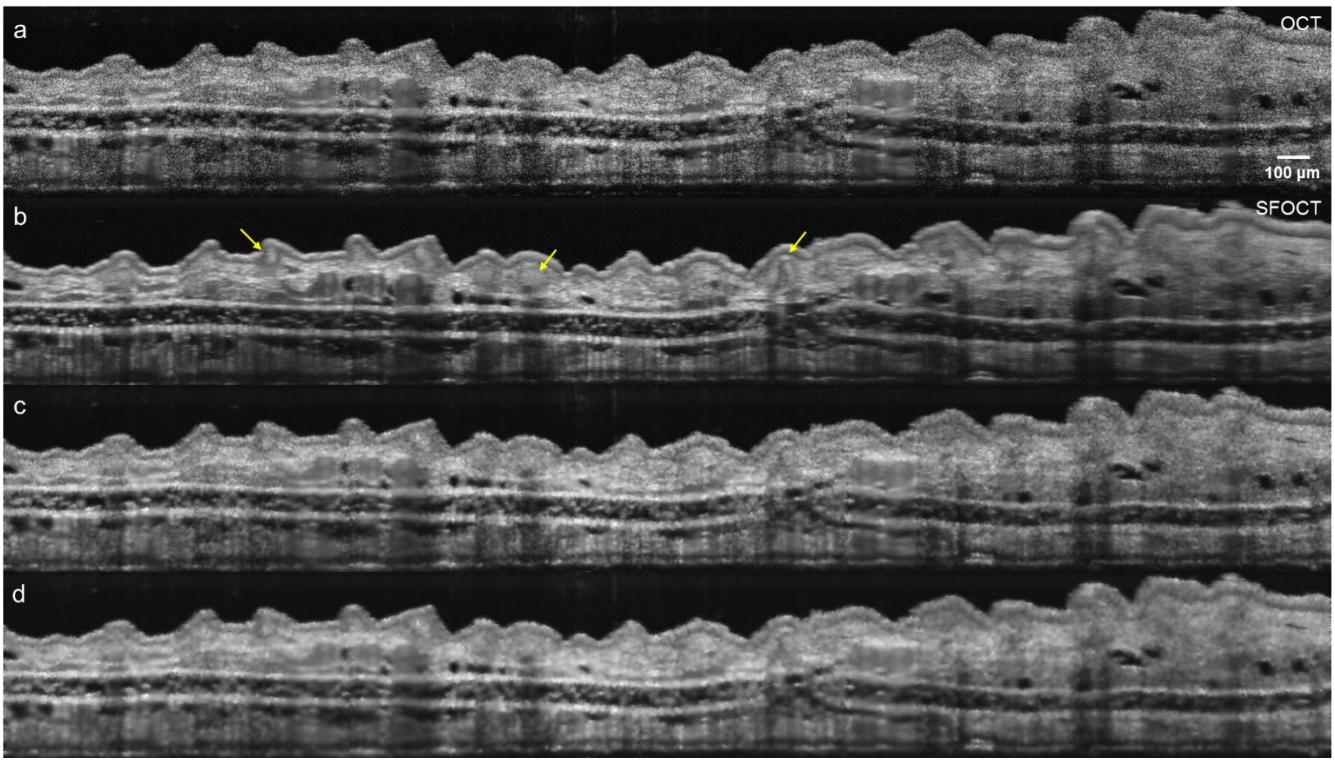

**Figure S12| Comparison of SFOCT to 2D digital filtering methods for speckle removal. a,** OCT image of a mouse pinna. The pixel size is 2x2 μm. **b,** SFOCT image at a similar location. Yellow arrows point to various fine structures. **c, d,** The OCT image post-processed with an adaptive Wiener filter using neighborhoods of sizes 5x5 and 7x7 pixels, respectively.

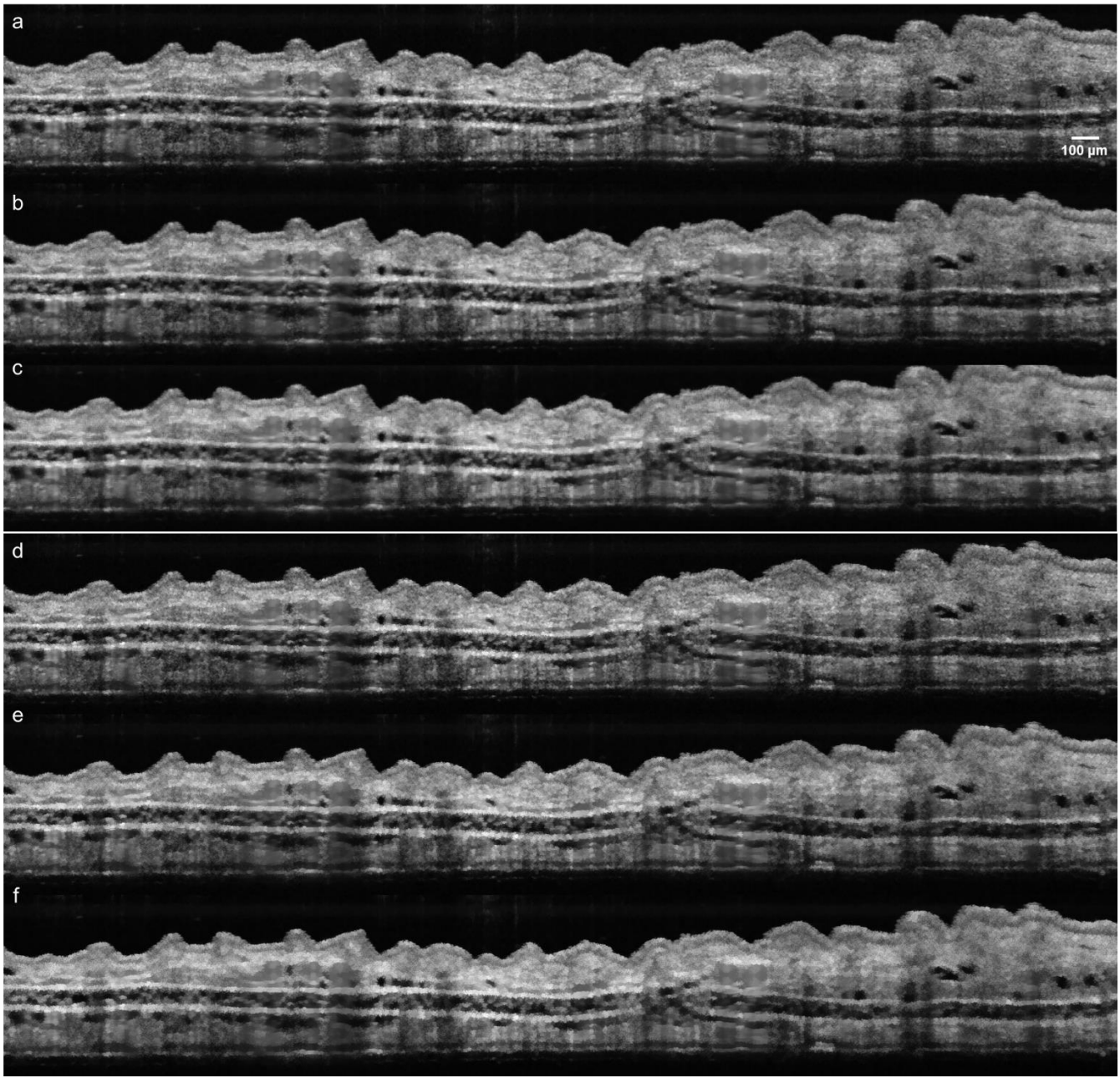

**Figure S13| Comparison of SFOCT to 2D digital filtering methods for speckle removal (continued). a-c,** The OCT image post-processed with a hybrid median filter (HMF) using neighborhoods of sizes 5x5, 7x7, and 9x9 pixels, respectively. **d-f,** The OCT image post-processed with a symmetric nearest-neighbor filter (SNN) using neighborhoods of sizes 5x5, 7x7, and 9x9 pixels, respectively. These examples show that filters of a small size do not reduce speckle noise sufficiently, whereas, larger filters create artifacts and compromise some of the information in the image.

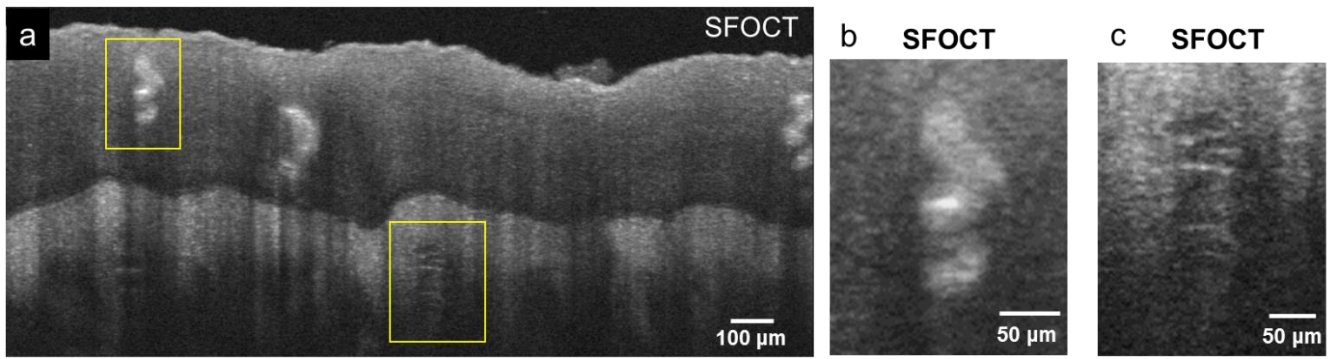

**Figure S14| SFOCT image of intact human fingertip skin. The acquisition parameters are similar to the OCT image shown in Fig. 5 a-c. a,** SFOCT B-scan of a fingertip. **b,** Close-up view on the sweat duct marked in **a**. **c,** Close-up view on the tactile corpuscle marked in **a**.

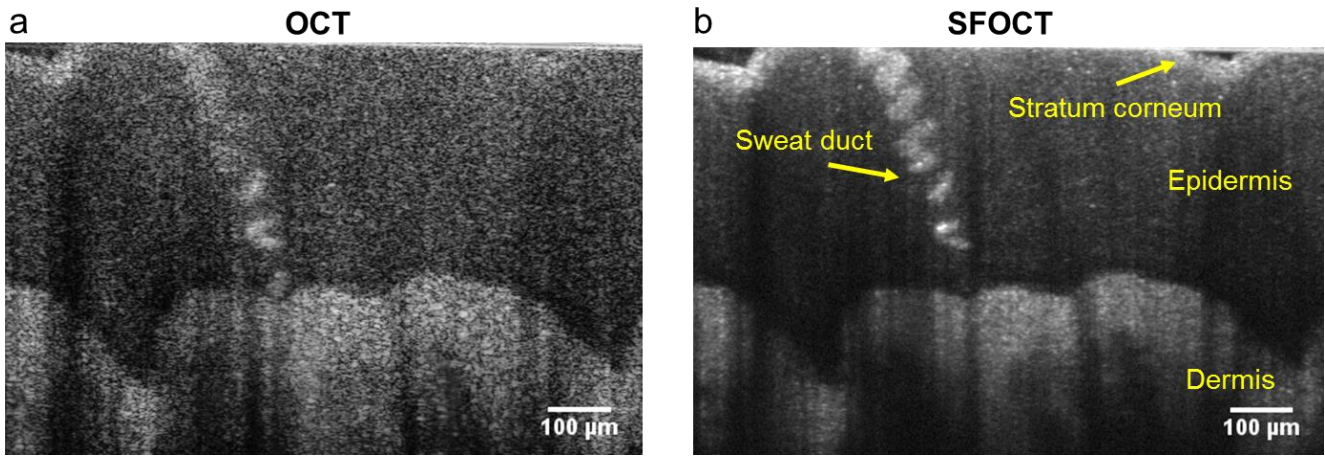

**Figure S15| OCT and SFOCT images of intact human fingertip skin. a,** OCT B-scan of a fingertip showing a sweat duct. **b,** SFOCT B-scan of a similar region, showing the sweat duct in greater detail, along with a better view of the layers in the skin.

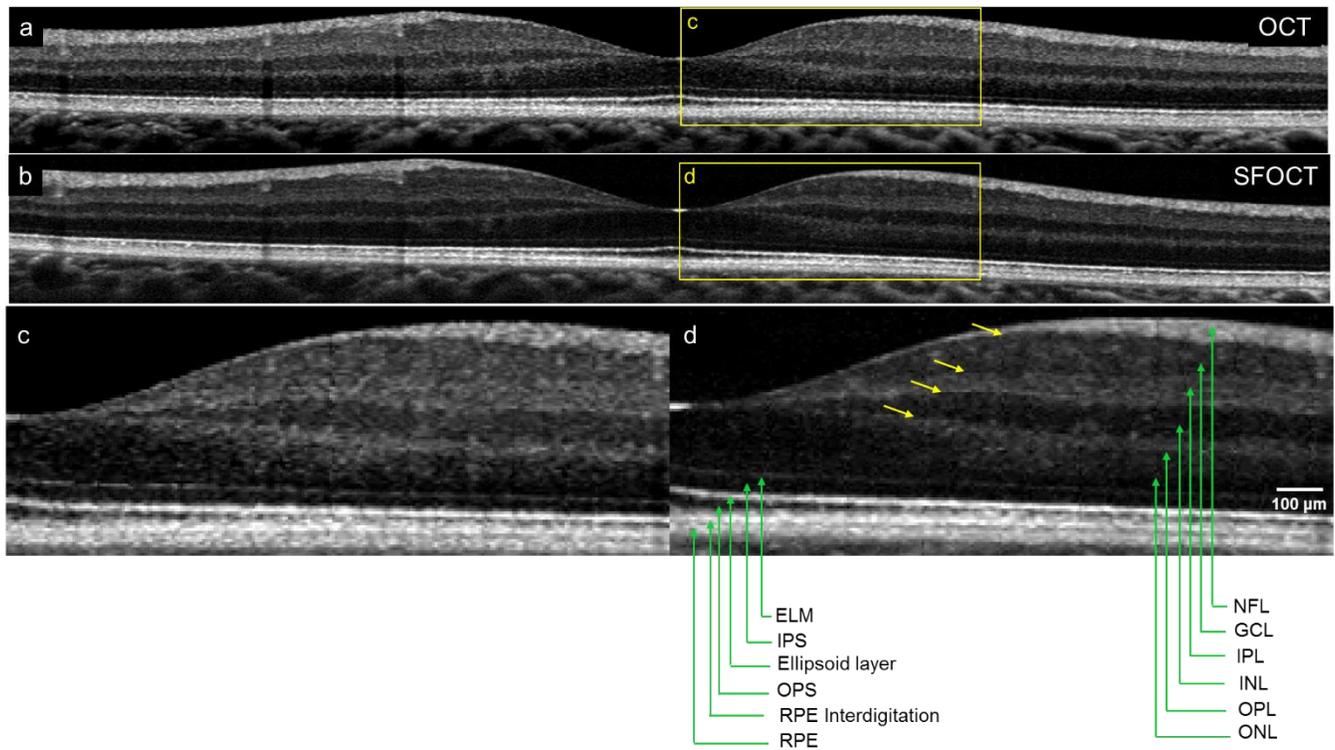

**Figure S16| SFOCT imaging of human retina shows clearer differentiation between the retinal layers without loss of detail. a,** In the top panel, the normal human retina is seen through a B-scan (cross-section) of the fovea using commercially-available spectral domain OCT (SDOCT). **b,** The same scan through the same human retina is shown using SFOCT (middle panel). **c,** Bottom left panel depicts magnification of the yellow box in **a** with SDOCT. **d,** Bottom right panel depicts the corresponding region magnified in **b** using SFOCT. Optical removal of speckle results in a clearer delineation of the various retinal layers, seen most clearly in the interfaces between the nuclear layers and the adjacent plexiform layers and/or nerve fiber layers (yellow arrows), the ellipsoid layer, OPS, RPE Interdigitation, and RPE. ELM=external limiting membrane, IPS=Inner photoreceptor segments, OPS=Outer photoreceptor segments, RPE=retinal pigment epithelium, NFL=nerve fiber layer, GCL=ganglion cell layer, IPL=inner plexiform layer, INL=inner nuclear layer, OPL=outer plexiform layer, ONL=outer nuclear layer.

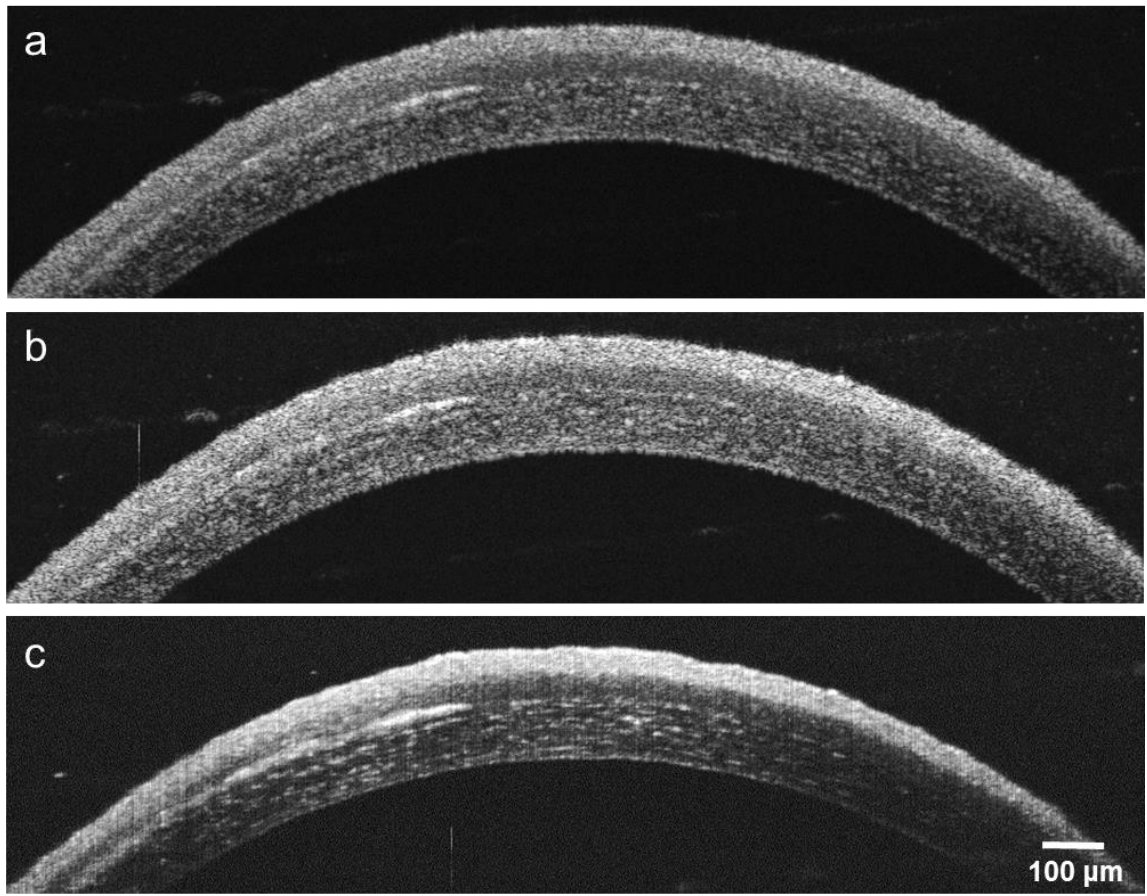

**Figure S17| OCT and SFOCT images of a mouse cornea with A-scan averaging instead of frame averaging. a,** OCT image composed by averaging 100 frames (B-scans). **b,** OCT image composed by averaging 100 consecutive A-scans. **c,** SFOCT image composed by averaging 100 consecutive A-scans. The speckle pattern is uncorrelated between A-scans owing to a fast moving (rotating) diffuser. The SFOCT image shows the lamellar structure of the cornea, which is hidden by speckle noise in the OCT images. The left side of the cornea shows increased light scattering of unknown origin.

# Tables

| Figure | Frame averaging | Pixel spacing x [μm] | Pixel spacing y [μm] | Diffuser used |
|---|---|---|---|---|
| 1f, h | 30 | 2 | 2 | - |
| 1g, i | 30 | 2 | 2 | 1500 |
| 2 f, h | 100 | 2 | - | - |
| 2i | 5, 10, 20, 40, 100 | 2 | - | - |
| 2 g, j | 100 | 2 | - | 1500 |
| 2k | 5, 10, 20, 40, 100 | 2 | - | 1500 |
| 3 a, b, e | 100 | 2 | - | - |
| 3 c, d, f | 100 | 2 | - | 1500 |
| 3g | 20 | 4 | 4 | - |
| 3h | 20 | 4 | 4 | 1500 |
| 4 a, b, f, g | 69* | 2 | - | - |
| 4 c, d, h, i | 81* | 2 | - | 1500 |
| 5 a, b, c | 20 | 4 | 4 | - |
| 5 d, e, f | 100 | 2 | - | 2000 |
| S5 a, d | 54 | 3 | 3 | - |
| S5 b, e | 54 | 3 | 3 | 2000 |
| S5 c, f | 54 | 3 | 3 | 1500 |
| S6 b | 30 | 2 | 2 | - |
| S6 c | 30 | 2 | 2 | 2000 |
| S6 d | 30 | 2 | 2 | 1500 |
| S8 a | 100 | 2 | - | - |
| S8 b | 100 | 2 | - | 2000 |
| S8 c | 100 | 2 | - | 1500 |
| S10 b | 100 | 2 | - | - |
| S10 c | 100 | 2 | - | 1500 |
| S11 | See caption | | | |
| S12 a, c, d | 100 | 2 | - | - |
| S12 b | 100 | 2 | - | 1500 |
| S13 | 100 | 2 | - | - |
| S14 | 20 | 4 | 4 | 2000 |
| S15 a | 100 | 2 | - | - |
| S15 b | 100 | 2 | - | 1500 |
| S16 a, c | 44 | ? | - | - |
| S16 b, d | 44 | ? | - | 3 μm |
| S17 a | 100 | 2 | - | - |
| S17 b | 100 A-scan averages | 2 | - | - |
| S17 c | 100 A-scan averages | 2 | - | 1500 |

* 100 frames were acquired. Several frames were removed due to movement of the sample.

**Table S1.** A full description of the acquisition parameters for all of the images in this study. A blank entry for pixel spacing in the y axis refers to a 2D scan, as opposed to a 3D scan. A blank entry for the diffuser used refers to a conventional OCT scan, as opposed to a SFOCT scan.

|  | Mean [μW] | STD [μW] | Relative transmission | Power loss | Diffuser transmission | Diffuser power loss |
|---|---|---|---|---|---|---|
| OCT | 814.75 | 4.24 |  |  |  |  |
| No diffuser | 738.25 | 2.61 | 91% | 9% |  |  |
| 3 μm lapped | 574.85 | 1.83 | 71% | 29% | 78% | 22% |
| 2000 grit | 593.55 | 1.9 | 73% | 27% | 80% | 20% |
| 1500 grit | 555.7 | 2.05 | 68% | 32% | 75% | 25% |

**Table S2.** Power on sample with conventional OCT and with the different diffusers.

|  | mean signal intensity [au] | STD [au] | Signal loss |
|---|---|---|---|
| no diffuser | 3.559 | 1.238 |  |
| 3 μm lapped | 2.285 | 0.812 | 36% |
| 2000 grit | 2.075 | 0.598 | 42% |
| 1500 grit | 1.769 | 0.486 | 50% |

**Table S3.** Signal intensity with conventional OCT and with the different diffusers.

|  | Smallest resolvable group (element) along vertical | Line pairs per mm | Effective FWHM [um] | Smallest resolvable group (element) along horizontal | Line pairs per mm | Effective FWHM [um] | PSF size increase |
|---|---|---|---|---|---|---|---|
| OCT | 6 (6) | 114 | 8.8 | 7 (3) | 161 | 6.2 |  |
| SFOCT, 2000 | 6 (6) | 114 | 8.8 | 7(2) | 144 | 6.9 | 6.18% |
| SFOCT, 1500 | 6 (1) | 64 | 15.6 | 6(6) | 114 | 8.8 | 35.27% |

**Table S4.** The effective lateral resolution of OCT and SFOCT, calculated from the visibility of line separation in the resolution target.

# Reference list for the Supplementary Information